\documentclass[useAMS,usenatbib]{mn2e}

\usepackage{natbib}
\usepackage{color}
\usepackage{amsmath, amssymb}
\usepackage{bm}
\usepackage{pifont}
\usepackage{float}
\usepackage{graphicx}
\usepackage[usenames,dvipsnames]{xcolor}
 \usepackage{relsize}
\usepackage{subfigure}  
\usepackage[singlelinecheck=false 
]{caption}

\usepackage{soul} 
\usepackage{epsfig}
\usepackage{enumerate}
\usepackage[shortlabels]{enumitem}
\usepackage[version=3]{mhchem}
\usepackage{caption}
\usepackage{varwidth}
\newsavebox\CBox
\usepackage{color}
\usepackage{lineno}
\usepackage{epsf,epsfig}
\usepackage{epstopdf} 
\usepackage{float}
\usepackage{rotating} 
\usepackage{epstopdf} 

     \def\d{\rm d}

\title[Flash ionisation signature in coherent cyclotron emission from Brown Dwarfs]{Flash ionisation signature in coherent cyclotron emission from Brown Dwarfs}
\author[I. Vorgul, Ch. Helling]{I. Vorgul$^{1}$\thanks{E-mail: iv4@st-andrews.ac.uk}, Ch. Helling$^{1}$\\
$^{1}$SUPA, School of Physics and Astronomy, University of St Andrews, St Andrews KY16 9SS, UK}

\begin{document}
\maketitle

\label{firstpage}

\begin{abstract}
Brown dwarfs form mineral clouds in their atmospheres, where charged particles can produce large-scale discharges in form of lightning resulting in substantial sudden increase of local ionisation. Brown dwarfs are observed to emit cyclotron radio emission. We show that signatures of strong transient atmospheric ionisation events (flash ionisation) can be imprinted on a pre-existing  radiation. Detection of such flash ionisation events will open  investigations into the ionisation state and atmospheric  dynamics. Such events can also result from explosion shock waves, material outbursts or (volcanic) eruptions. We present an analytical model that describes the modulation of a  pre-existing electromagnetic radiation by a time-dependent  (flash) conductivity that is characteristic for flash ionisation events like  lightning. Our conductivity model reproduces the conductivity  function derived from observations of Terrestrial Gamma Ray  Flashes, and is applicable to astrophysical objects with  strong temporal variations in the local ionization, as in planetary  atmospheres and protoplanetary disks. We show that the field  responds with a characteristic flash-shaped pulse to a conductivity  flash of intermediate intensity. More powerful ionisation events  result in smaller variations of the initial radiation, or in its  damping. We show that the characteristic damping of the response  field for high-power initial radiation carries information about the  ionisation flash magnitude and duration. The duration of the pulse  amplification or the damping is consistently shorter for larger  conductivity variations and can be used to evaluate the intensity of  the flash ionisation. Our work suggests that cyclotron emission  could be probe signals for electrification processes inside BD  atmosphere.
\end{abstract}




\section{Introduction}\label{sec:1}

 The search for lightning in the solar system has benefited from an
 increasing variety of space missions (e.g. Venus Express,
 Cassini). Lightning is the most observed flash transient process,
 that is a powerful but brief, short term processes leading to a local
 burst of ionisation.

Characteristic optical and radio emission provide growing evidence
for lightning in solar-system planets \citep{rakov07,farrell99}.
Powerful flash transient processes can act destructively, or they can
trigger the formation of new (e.g. pre-biotic)
molecules~\citep{miller59}. On Earth, lightning is one of the most
significant sources of natural ozone production
e.g.~\citep{sangh08}. Associated with terrestrial lightning is
powerful gamma ray emission (Terrestrial Gamma ray Flashes (TGFs)\footnote{TGFs are suggested to be produced by runaway
  electrons, and lightning is suggested to be a source of the
  electrons' acceleration \citet{ostgaard13,maris13}.}; 
e.g.,~\citealt{fishman94,dwyer12,carlson10}).

 We are particularly interested in lighting as an example for flash
 transient processes in atmospheres of Brown Dwarfs and Giant Gas
 Planets with respect to observations of cyclotron maser emission that
 is observed in increasing number of brown dwarfs
 ~\citep{hall07,yu12,will14,will13}. Brown dwarfs have clouds and
 weather-like variation that determines their dynamic
 atmospheres~\citep{metchev13,cross,buenzli,radigan,apai}. Recent
 Spitzer observations suggest the detection of global
 winds~\citep{heinze13}. \cite{hel13a} suggest that in such clouds
 small-scale spark discharge can be expected at cloud bottom while
 large-scale discharge should be more common near the top of the cloud
 or above it. Hence, flash ionisation processes can be expected to
 occur rather frequently in the cloud layers of these ultra-cool
 objects. \cite{bailey14} show that such flash ionisation events may
 affect a larger volume of the atmosphere in Brown Dwarfs than in the
 atmosphere of Earth. 

The observation of cyclotron emission from brown dwarfs confirms
  that strong magnetic fields are present in these ultra-cool objects
  \citep{hall07,berger10,burgasser13,yu12,will13,will14}. The
  observation of a strong and coherent cyclotron emission further
  confirms an existence of a source of ionisation to produce the electron beams responsible for the emission.

For hotter stars than M-dwarfs and brown dwarfs, coronal mass
 ejections are the sources of accelerated electrons like for example on
 the radio emitting flare stars like UV Ceti or the magnetic chemically
 peculiar, extremely fast rotating radio pulsar CU
 Virginis~\citep{kellett01,vor11}. The solar wind provides the
 majority of electrons for planetary cyclotron radio
 emission~\citep{zarka92} for most of the solar system planets\footnote{Jupiter is slightly off-set from the straight line dependence of solar system planets' radio emission on the solar wind flux, and sources of the accelerated electrons there, including Io volcanic activity, are still debated.}.  Most brown
 dwarfs, however, occur as single stars or in brown dwarf - brown
 dwarf/ultra-low mass binary systems. Non of the above sources of free

 electrons is, hence, available to brown dwarfs, unless they are part
 of a binary system where an active M-dwarf producing coronal mass
 ejections or a white dwarf provides high-energy irradiation
 (\citealt{Casewell2015}). Cyclotron emission from brown dwarfs is
 powerful and coherent, and the observations of this emission are
 consistent and reproducible. Possible sources of
 free electrons leading to strong radio emissions in ultra-cool
 objects like brown dwarfs and free-floating planets are under
 investigation and include dust-dust collisions~\citep{hel12,hel13a},
 Alfven ionisation~\citep{stark13} and cosmic rays
 impact~\citep{rim13}. Electric discharges (lightning, transient
 luminescent events, small-scale but frequent coronal discharges) are
 additional sources for electrons.

Given that brown dwarfs are powerful radio emitters and that
  clouds and winds form in their atmospheres, we suggest using
  cyclotron emission as a probe signal (carrier signal) to search for
  its transformation by atmospheric processes. Recently,
  \cite{Schellart2015} observed that the lighting radio emission
  originating from relativistic electrons accelerated in the Earth
  magnetic field imprint their signal onto the radio signal of cosmic
  ray induced atmospheric air showers. In \cite{Schellart2015}, the
  carrier signal for the effect of lighting is the radio emission from
  a cosmic ray airshower, while a more powerful source is required
  for the carrier source in the astrophysical context.

Existing data from brown dwarfs may already carry the fingerprints of lightning 

in a form of specific time-variations.  By studying how a coherent
(cyclotron) emission is modulated\footnote{\textit{Modulation of the
    carrier frequency} refers to the energy transfer between the field
  and the atmospheric gas that undergoes a flash ionisation, and not
  the field modulation by another radiation field} by flash transient
 processes, we suggest a new detection method for transient events
  and for the atmospheres where they occur.  We demonstrate how
  the electromagnetic field of the cyclotron radiation is transformed
  and how the signatures of the flash ionisation is imprinted onto the
  pre-existing cyclotron emission.  \textit{Flash ionisation}
describes the fast transient events leading to rapid ionization of
surrounding medium lightning being one example~\citep{hel12}. Other
discharges include those induced by cosmic rays~\citep{gur96},
explosions, plasma jets or lightning in volcano plumes.

\bigskip


Section 2 contains our approach with details of electron cylotron
emission and a reflection of emission symmetries. The concept of the
time-dependent conductivity as a model for lightning is outlined in
Sect. 3. Section~\ref{sec:2} presents our mathematical modelling for
the electromagnetic field transformed by a flash ionisation process
represented by a time-dependent conductivity. An exact analytical
solution to the equations for a general conductivity function is given
in Sect.~\ref{ssec:2.2}. The temporal variation of the particular
flash-like conductivity is modelled in
Sect.~\ref{ssec:2.3}. Section~\ref{ssec:3.4} presents a test case for
our model by comparing the conductivity time-variations with TGF
observations. The formal body derived in Sect~\ref{sec:2} is not
restricted to radio waves only.  Section~\ref{sec:3} presents a
parameter study that shows how the electromagnetic field responds to
different conductivity time-flashes. We demonstrate how the amplitude
and the damping character changed depending on the strength of the
intersected flash ionisation event. Section~\ref{sec:4} discusses the
observational effects of flash ionisation events on radio emission
from Brown Dwarfs. We summarise this section in a 'Recipe for
observations'.

\section{Approach}\label{sec:7}

This section outlines the approach that we take to derive the
signatures of flash ionisation events like lighting imprinted onto a
pre-existing cyclotron emission. We first summarise necessary facts
about electron cyclotron maser emission (Sect.~\ref{ss:CE}) and
outline our approach to cyclotron emission probing lightning-active
regions in Sect~\ref{ss:how}.

\subsection{Cyclotron emission}\label{ss:CE}

\begin{figure*}
	\centering\includegraphics[width=0.75\textwidth]{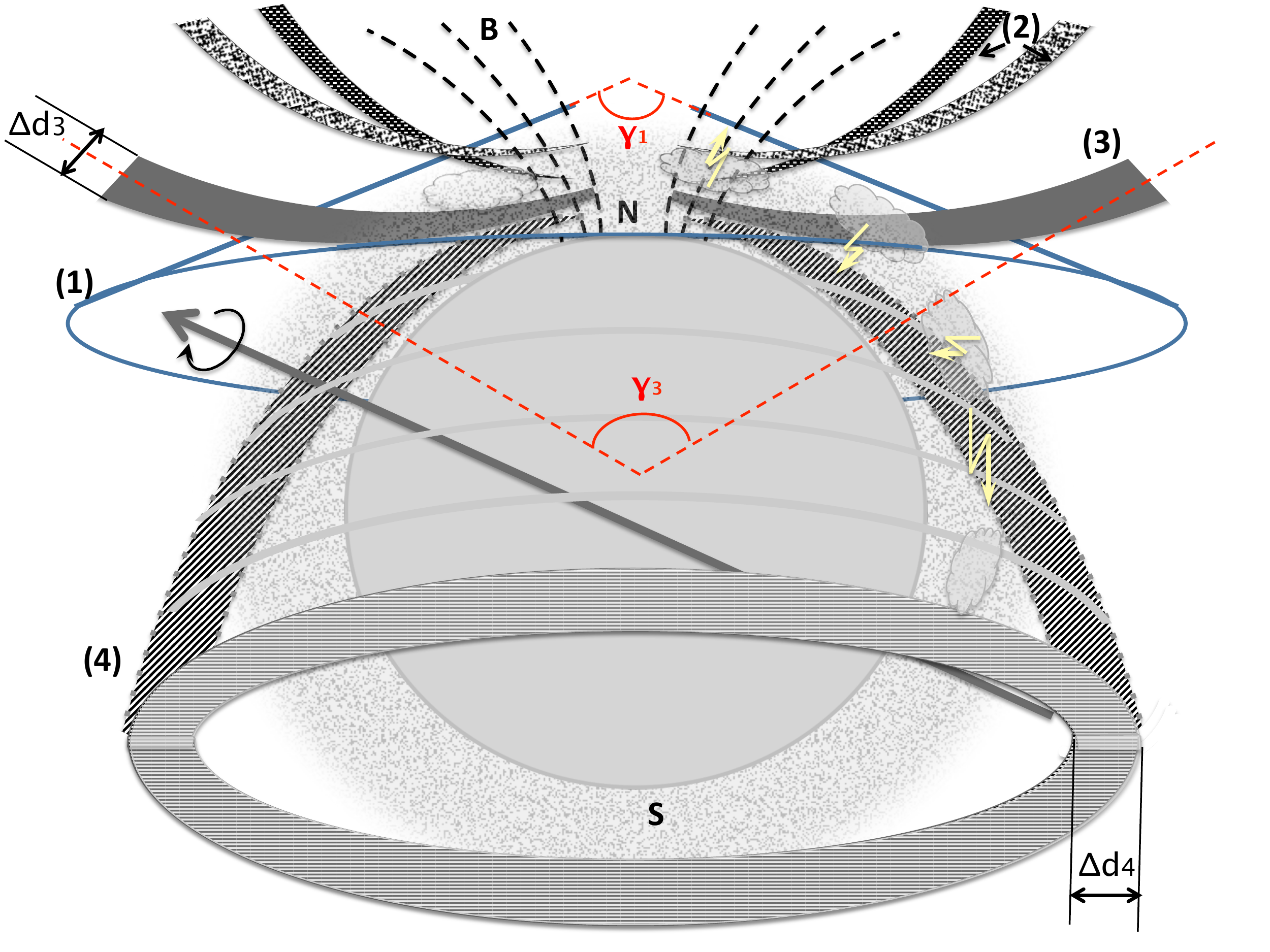}
	\caption{Schematic representation of the radiation cones
          formed by electron cyclotron emission that originates from a
          magnetic pole and atmospheric interaction area. Depending on
          the source area at the magnetic field line near the magnetic
          pole, the emission cones, (1)$\ldots$(4), have different
          opening angle ($\gamma_1\ldots\gamma_4$) and are refracted
          differently through the atmosphere. The path taken by the
          radiation will determine the wall thickness ($\Delta
          d_1\ldots\Delta d_4$) with $\Delta d_4$ being the thickest of
          the columns of refracting material passed by the radiation. The
          magnetic pole (top) does not coincide with the rotational
          axis.}
	\label{f:11}
\end{figure*}

Electron cyclotron emission is a localised-source emission, which is known to
emerge out of a star/planet's atmosphere shaped as a hollow cone
centred around a magnetic pole (e.g.,~\citealt{bingham13}). Each
  cone is associated with a distinct source region as indicated in
  Fig~\ref{f:11}. Such radiation cones form due to a high directivity
property of the local cyclotron emission (e.g.,~\citealt{speirs13}),
with its annular symmetry reflecting the magnetic field symmetry around
the pole. Rotation of the object causes the periodicity of the
radiation peaks detected when the cone's walls cross the receiver's
line-of-sight. The width of the observed radiation peaks
  (e.g. Fig. 1 in \citealt{Hallinan2008}) is
  determined by the extension of the source and by the path length
  along which the radiation propagated and along which it experiences
  scattering and dispersion effects. It is therefore proportional to
  the cone-wall thickness, $\Delta d$, as depicted in Figs.~\ref{f:11}
  and~\ref{f:1}.

 Cyclotron maser emission occurs when the velocity distribution
 function of the electron beam, that travels into increasing magnetic field,
 is continually transformed by magnetic compression to form a
 horseshoe-shape distribution in velocity space\citep{kellett01}. When eventually the
 horseshoe is confined enough for the major electron population
 to satisfy the cyclotron resonance condition,
 $\omega=\frac{\omega_\text{c}}{\gamma}-k_{\parallel}
 v_{\parallel}$\,(\footnote{ where $\omega_\text{c}$ is a local cyclotron
   frequency, $\gamma$ is a relativistic factor,
   $\gamma=\left(1-\frac{v_{\perp}^2+v_{\parallel}^2}{c^2}\right)^{-1/2}$,
   $c$ is the speed of light, and $k_{\parallel}$ and $v_{\parallel}$
   are wave number and particle velocity parallel to the beam
   direction, respectively}), the beam emits the cyclotron maser
 radiation at the location where this resonance happened. 
 Various factors influence the atmospheric altitude at which this resonance
 occurs, including the initial electrons distribution in velocity space, the location where
 the electrons coupled to the magnetic field, and the magnetic
 field values, gradient and topology. 

It is known from Earth and
Saturn direct in-source observations (e.g., \citet{ergun00, lamy10}
to happen inside plasma cavities with rarefied plasma. This does not
imply the average local density should be low, while higher density
can influence the radiation escape. Being emitted from a localised region with small spread along the altitude, the radiation\footnote{from a consistent and parameter-stable beam} at the source has a sharp frequency spectrum. In addition to the in-situ observations, this was confirmed by scaled laboratory experiment~\citep{sandra08}. The emission frequency therefore corresponds to the local magnetic field and depends on the altitude at which the emission happened.

 The cyclotron radiation is emitted nearly perpendicular to the
 magnetic field lines (e.g., \citet{mutel08, speirs13escape}), but it
 can be refracted by the atmospheric gas and clouds before it emerges from
 the atmosphere in the shape of a hollow cone (Fig.\ref{f:11}). This
 can also be caused by the non-uniform magnetic field along the path and,
 consequently, non-uniform plasma frequency and non-uniform refractive
 index~\footnote{Refractive index of plasma in presence of magnetic
   field depends on plasma frequency and on the magnetic field, in
   particular on the angle between the wave vector and the local
   magnetic field, and is anisotropic in general case. When the
   magnetic field has its direction and magnitude changing along the
   wave propagation path, the refractive index for this wave will be
   inhomogeneous. }.  If the source is far enough from the ionosphere
 and the radiation is not influenced by propagation effects (case 1 in
 Fig.\ref{f:11}), the radiation will propagate in a very wide straight
 cone (almost plane-like, hence $\gamma_1$ very close to $180^o$) with
 very thin cone walls (opening angles $\gamma_1\gg \gamma_3$, hence
 cone wall thickness $\Delta d_1 \ll \Delta d_3$ in
 Fig.~\ref{f:11}). Radio pulsars are examples of the sources of such straight (not refracted) propagation. Typical width of the radio emission cones for pulsars, $\Delta d$, is around $2$--$8^o$ (e.g., $7.22^o$, $5.8^o$ and $4.2^o$ found in \citet{kramer94}).  A summary of the underlying physics and geometry consideration addressing the question why the cyclotron
 emission appears in hollow cones is given in \cite{depater10}. Radio
 observations from brown dwarfs, however, often suggest smaller cone opening
 angles, $\gamma$, and thicker cone walls ($>10^o$), $\Delta d$, due to
 broadened (by propagation) radiation pulses in the observed phase curves. The pulse
 broadening would be caused by interaction of the emitted radiation
 with the atmospheric gas and clouds, hence, larger broadening
 advocating a larger path lengths (case 3 or 4 in Fig.~\ref{f:11}) in
 contrast to very narrow pulses of non-refracted radiation
 (case 1 in Fig.~\ref{f:11}).  Higher frequency cyclotron radiation from the same astrophysical source is emitted deeper into the magnetic field/atmosphere (closer to the magnetic
 pole where the magnetic field lines converge; see cases 3 and 4 in
 Fig.~\ref{f:11}), and hence it would
 refract steeper than lower frequency one.  
 
 In this case, it is likely to pass through more of the
 stratified atmospheric volume becoming more affected by the propagational effects, which results in thicker
 walls of the radiation cones. Note that the effect of different refraction of different frequencies' cones could only be noticeable in observations for a substantial separation of the frequencies to allow for a substantial separation of the local medium properties. E.g., with the cyclotron frequency proportional to the local magnetic field ($\nu_{\rm e}=e\cdot B/(2\pi m_{\rm e} c)$), which decreases from
 the star/planet's surface as $\frac{1}{R^3}$ ($R$ being a radial distance from the object's magnetic dipole centre), the  square of the refractive index determined as   $n^2=1-\frac{\nu_p}{\nu (\nu \pm \nu_e )}$ (where $\nu_p$ is the local plasma frequency and $\nu$ is the frequency of the radiation) will also have roughly $\frac{1}{R^3}$ dependence. Therefore, the difference between $n^2$ at different locations along the magnetic field lines would be proportional to $\frac{1}{R_1^3}-\frac{1}{(R_1+\Delta R)^3}\eqsim 3\cdot \frac{\Delta R}{R_1^4}$, which means that the relative difference in refraction index, $\frac{\Delta n^2}{n_1^2}$, is $3\cdot \frac{\Delta R}{R_1}$ times smaller than the correspondent relative frequencies difference, $\frac{\Delta \nu_e}{\nu_\text{e1}}$. 
 
The effect of two well-separated (in frequencies) cones refracting differently on their way out of the star's magnetosphere is significant in observations from CU Virginis. The effect of refraction was suggested as an explanation
for multi-frequency observations to explain why
CU Virginis radio observations (e.g., (Kellett et al. (2007); Trigilio
et al. (2008))) found a higher frequency cone (i.e., radiated
deeper into the magnetic field and the atmosphere, at lower
altitude) inside a lower frequency one, radiated at higher latitude
(Lo et al. 2012)). The clearly seen refraction there resulted in the cones' walls, $\Delta d$, broadening up to $13.80^o$ and $14.48^o$ (Kellett et al. (2007)).

 The idea of cyclotron emission radiated in the shape of hollow cones
 was applied by \cite{Wang1995} to derive an opening angle of
 $\gamma=80^o$ and a cone wall thickness $\Delta d = 15^o$ for
 Neptune's radio emission, and \cite{Imai2002} applies the concept to
 Jovian radio emission.  Observations by \cite{doyle10} for the
 brown dwarf TVLM 513 suggest an estimation of the opening angle from their Fig.3 being $\gamma = 0.17\cdot
 360^o=61.2^o$, and the observed FWHH of the radiation peak in the same plot can provide an estimation of the
 radiation cone wall thickness as $\Delta d=0.065\cdot
 360^o=23.4^o$. Observations by \cite{lynch15} present the phase curve of the emission (their Fig. 4) from which the cone wall thickness can be estimated as $\Delta
 d=1^o\ldots 2^o$ ('beam angle' in \citealt{lynch15}). This small
 $\Delta d$ may indicate the cyclotron source located rather high in
 the atmosphere, comparable to case 2 in Fig~\ref{f:11}.
 \cite{will15-multiwaveUCD}'s observation for NLTT 33370AB (their
 Figs. 3, 4; rotation period P=3.7h) could be interpreted as a more complicated
 scenario with a rather small radiation cone with an opening angle estimated from the phase curve as
 $\gamma=35^o$ and a wall thickness of $\Delta d=20^o\ldots23^o$. Such
 wide beams (large $\Delta d$) (comparable to the refracted CU-Virginis beams while being broader than non-refracted pulsar ones) with a relatively small opening angle
 ($\gamma$) may be represented by case 4 in Fig~\ref{f:11}.  We note
 that the number of cones depends on the magnetic field geometry as
 demonstrated in \citet{kellett07}.

\subsection{How can cyclotron emission probe an atmosphere with lightning-active regions?}\label{ss:how}

 Brown dwarf atmospheres and similarly cold atmospheres, like those of
 extrasolar giant gas planets, are made of chemically diverse
 oxygen-rich, H$_2$-dominated gases. Clouds form and have a distinct
 influence on the atmosphere through their opacity and through element
 depletion. Brown dwarf atmospheres can also be expected to be dynamic
 as brown dwarfs are fast rotators. Giant gas planets may be close to
 their host star such that irradiation drives strong winds inside
 their atmospheres. It is plausible to expect that such dynamic, cloud
 forming atmospheres generate lighting activity (\citealt{hel12,
   Helling2013_767}) more intensive than in the solar system and
 affecting larger atmospheric volumes (\citealt{bailey14}). 

 Ultra-cool atmospheres may show different kind of discharges in form
 of thunderstorms including inter-cloud and intra-cloud discharges.
 Transient luminous events (sprites, elves, jets, etc.) are frequently
 observed on Earth in conjunction with lighting.  Sprites and
 jets are substantially more extended than lighting, and therefore
 increase the atmospheric volume affected by flash ionisation
 events. These flash ionisation events perturb the upper atmosphere and
 magnetosphere (e.g.  \citealt{siingh09}).  For a brown dwarf with a
 radius of $\sim 7\cdot 10^4$ km and a cyclotron peak's half-width of
 about $\delta \theta =2^o$ \citep{kuznetsov12,hall07}, the spread of
 the beam ($\Delta d$ in Fig~\ref{f:11}) emerging from the atmosphere
 is $\Delta d=2\pi\cdot 7\cdot 10^4$km$ \cdot 2\deg/369\deg = 2444$\,km.
 If this radiation beam underwent a ten-fold spatial dispersion
 (for example case 3 in Fig.~\ref{f:11}), the source region near the
 magnetic pole would have an extension (i.e., the cone's wall thickness) about 244 km across. This horizontal (with respect to the star/planet's surface) extend  of a storm between $244$ km and $2444$ km determines the volume of the column of the atmosphere which could be probed by the cyclotron beam. Recent observation from Saturn's
 storm area indicate that the storm that emerges at the surface alone
 has a head diameter of $\sim 2000$km. \cite{Zhang2014} demonstrate
 how the atmospheric dynamic for brown dwarfs produces similar
 pattern. These rough ideas of the radiation/storm longitudinal extensions suggest them to be reasonably comparable for the storms being able to affect a significant portion of the beam's radiation. 
 We suggest that if the ECM radiation travels through the extended atmosphere of the brown dwarf/planet, it may be modulated by conductivity changes in the atmosphere produced by flash ionisation and can proposed to probe the
 extended atmosphere it travels through.

In the following, we consider a local fragment of the radiation cone
at a direction that will rotate into the observer's
view. Cyclotron emission is a frequency-resonant phenomenon and can
therefore be represented by a harmonic wave. We study how this
coherent wave changes after experiencing a flash ionisation event
affecting a volume it paths through.

\section{Analytical model for the electromagnetic field transformation caused by time-dependent ionisation event}\label{sec:2}


We consider a pre-existing coherent radiation (cyclotron radiation or
similar) travelling through an atmosphere environment with
flash-ionisation events (e.g. lighting, sprites, flares).  Its
coherent character allows to model this incident electromagnetic field
as a harmonic wave. We are interested in how the electromagnetic field
of a pre-existing emission is transformed by the flash ionisation.

The interaction between an electromagnetic wave and a medium rapidly
changing in time is highly non-linear. Current approaches that
model radiation, emitted by flash ionization processes, are focused on
microscopic modelling of charged particle production (electrons and
ions; e.g.,~\citealt{carlson10,tsalkou13}), in some cases including
electric (\citealt{luque13}) or magnetic (\citealt{maclachlan09})
fields. As a result, microscopic approaches predict linear growth
rates of the number density of charged particles, and hence, the
linear growth stage of electromagnetic emission only.  In order to
determine the maximum of the flash ionisation event, its decline and
duration, the non-linear saturation and the consequent drop in the
particles velocities and production rate needs to be taken into
account. This will allow us to qualify and quantify the
electromagnetic field response.  In the case of an external field
probing the flash-ionisation area, non-linear interactions between
field and the transient, ionised medium can be expected to be stronger
compared to the discharge’s own radiation field because of the higher
amplitude of the initial (cyclotron) field.  We therefore apply the
concept of a parameterised time-dependent conductivity in a
macroscopic model to allow us to account for non-linear interactions
between the ionised medium and the field. We use Maxwell's equations
formalism to study the field transformation by the transient
processes.  We model the flash ionisation by a time-dependent
conductivity with a flash-like shape. A flash-shape like conductivity
function first rises describing the fast linear growth of the ionised
particles at a breakdown ionisation stage, then reaching saturation
and eventually declines decribing the ending of the ionisation process
(see Fig.\ref{f:4}). This time-dependent functional form is inspired
by experimental and observational results
(e.g.~\citealt{gardner84,farrell99,lu11,aleks9}). Flash ionisation
events like lightning, coronal discharges, energetic explosions or
eruptions, can be suitably described by a time-dependent conductivity
without expensive small-scale modelling, because suitable variation in
amplitude, duration, shape and the flash front's gradient can be
adopted.  Conductivity shapes derived from TGFs is a particular well
studied case which we use as test case in Sect.~\ref{ssec:3.4}.

In Sect.~\ref{ssec:2.1}, the changing electric field component from
Maxwell's equations for a time-dependent conductivity flash is
derived, and the exact solution of the derived equations
(Eqs.~\eqref{e:6}--\eqref{e:8}) is presented in
Sect.~\ref{ssec:2.2}. The time-dependent conductivity
function is discussed in Sect.~\ref{ssec:2.3} and its relvance of natural
discharges in Sect.~\ref{ssec:3.4}.

\subsection{Deriving the equation for the electric component of the electromagnetic field}\label{ssec:2.1}


The electromagnetic fields, $\vec{E}(t,\vec{r})$ and
$\vec{H}(t,\vec{r})$, that interact with a fast-changing medium
influence the medium parameters (i.e. dielectric permittivity or
conductivity), and these induced changes feed back into the fields. In
the model presented in this paper, the electromagnetic fields in a
transient medium are described by the linear Maxwell's
equations. However, the material coefficients, dielectric
permittivity, magnetic permeability or conductivity, as part of
Maxwell's equations in turn depend on the fields, and therefore the
field equations are intrinsically non-linear. This section presents
the derivation of the equations for the electric component of the
electromagnetic field for a time-dependent conductivity. The
interaction of the pre-existing coherent radiation field with the
transient event starts at the time $t=0$. At this moment, the medium's
conductivity begins changing in time in a flash-like manner due to the
ionisation process.

We consider the problem in one spacial dimension, $x$, and assume that the electromagnetic fields only have components which are perpendicular to that dimension,
\begin{equation}\label{e:1}
\begin{aligned}
&\vec{E}(t,\vec{r})\equiv E_z(t,x)\equiv E(t,x)\,, \\ 
&\vec{H}(t,\vec{r})\equiv H_y(t,x)\equiv H(t,x)\,.
\end{aligned}
\end{equation}

The fields should satisfy Maxwell's equations (in CGS units),
\begin{equation}\label{e:1a}
\tag{\ref{e:1}a}
\left\{%
\begin{aligned}
&\nabla \times \vec{E}=-\frac{1}{c}\frac{\partial\vec{H}}{\partial t},\\
&\nabla \times \vec{H}=\frac{4\pi\sigma(t)}{c}\vec{E} +\frac{\varepsilon}{c}\frac{\partial\vec{E}}{\partial t},
\end{aligned}
\right.
\end{equation}
where $\varepsilon$ is a dielectric permittivity of the medium, which
is dimensionless and is considered to be constant in time, $c$ is the
velocity of light [cm/s], $\sigma$ is the time-dependent conductivity
[1/s], and $E$ [statV/cm] and $H$ [Oe] are the electric and magnetic
components of the electromagnetic field, correspondingly. Considering
1D electric and magnetic field components only, as in Eq.~\eqref{e:1},
the following equation is derived from the system of two Maxwell's
equations for $E$ and $H$ components of the electromagnetic field
(Eq.~\eqref{e:1a}), by excluding $H$ and collecting the time-dependent
conductivity term in the right-hand side:
\begin{equation}\label{e:2}
\frac{\partial^2}{\partial x^2}E(t,x)+\frac{\varepsilon}{c^2}\frac{\partial^2}{\partial t^2}E(t,x)=-\frac{\partial}{\partial t}\biggl(\frac{4\pi\sigma(t)}{c}E(t,x)\biggr)\,.
\end{equation}
The left-hand side of Eq.~\eqref{e:2} is a wave equation for homogeneous media. We assume that the transient processes start at $t=0$. The left-hand side of Eq.~\eqref{e:2} has a known Green's function~\citep{khizhnyak86}, which is a solution of a wave equation with the Dirac's delta function $\delta(t-x/v)$ right-hand side (RHS) denoting a point source of radiation. Applying convolution of the Green's function to its right-hand side~\citep{ner12} the following integral equation for the field results~\citep{vor07,speirs13}:
\begin{multline}\label{e:3}
E(t,x)=E_0(t,x)-\\
{}\frac{2\pi}{\varepsilon v}\int\limits_0^\infty \d t'\!\!\int\limits_{-\infty}^{\infty} \d x' \sigma(t')\,\delta \biggl(t-t'-\frac{|x-x'|}{v}\biggr)E(t',x')\,,
\end{multline}
where $v=c/\sqrt{\varepsilon}$, and $E_0(t,x)$ is the initial electric component of the field, describing the pre-existing cyclotron radiation (i.e. the field which propagated through the medium before the transient processes starts at ${t=0}$).

We consider the initial field $E_0(t,x)$ as a plane wave, $E_0(t,x)=E_0 e^{-i\omega\left(t-\frac{x}{v}\right)}$, and assume the transformed field's spatial dependence in the similar form, as
\begin{equation}\label{e:4}
E(t,x)=E(t)e^{i\omega \frac{x}{v}}.
\end{equation}
Equation~\eqref{e:3} can then be reduced to the following equation for the field's time-dependence $E(t)$, after integrating Eq.~\eqref{e:3} over $x$,
\begin{equation}\label{e:5}
E(t)=E_0 e^{i\omega t}-\frac{4\pi}{\varepsilon}\int\limits_0^t \d t' \sigma(t') \cos\bigl(\omega(t-t')\bigr) E(t')\,.
\end{equation}

Equation~\eqref{e:5} can be transformed after double integration into a second-order differential equation for the field's time-dependence, $E(t)$,
\begin{equation}\label{e:5a}
\frac{\d^2 E(t)}{\d t^2}+\frac{4\pi\sigma(t)}{\varepsilon}\frac{\d E(t)}{\d t}+ \Biggl(\omega^2+\frac{4\pi}{\varepsilon} \frac{\d\sigma(t)}{\d t}\Biggr)E(t)=0\,.
\end{equation}
This is a linear differential equation, which accounts for nonlinear effects of the flash ionisation in Maxwell's equations, Eq.~\eqref{e:1a}, by introducing the time-dependent conductivity, $\sigma(t)$.

Substituting $t$ with a dimensionless $\tau$ by $t\to \tau=\omega t$, Eq.~\eqref{e:5a} takes a dimensionless form with respect to a normalised electric field, $E_1(\tau)=E(\tau)/E_0$
\begin{equation}\label{e:6}
\frac{\d^2 E_1(\tau)}{\d \tau^2}+f(\tau)\frac{\d E_1(\tau)}{\d \tau}+\biggl(1+\frac{\d f(\tau)}{\d\tau}\biggr)E_1(\tau)=0\,,
\end{equation}
where the function $f$ represents the conductivity time-dependence,
\begin{equation}\label{e:7}
f(\tau)=\frac{4\pi\sigma(t)}{\varepsilon\omega}\,.
\end{equation}
The initial conditions for $\tau$=0 are
\begin{equation}\label{e:8}
\begin{split}
E_1(0)&=1\,, \\
\frac{\d E_1(t)}{\d t}\biggr|_{t=0}&=i-\frac{4\pi\sigma(0)}{\varepsilon\omega}\,.
\end{split}
\end{equation}
Note, that the differential equation Eq.~\eqref{e:6} could be derived without using the Green's function and integral equation for the field, but using the integral equation allows finding the right initial condition for the field's first derivative, Eq.~\eqref{e:8}.

\subsection{Solution for the electric field component}\label{ssec:2.2}

We seek to solve Eqs.~\eqref{e:6}--\eqref{e:8} to study how an electromagnetic field in an initial form of a harmonic wave can be transformed by transient events. We aim to derive a signature of transients (like lightning) in a pre-radiated field such as coherent cyclotron radiation.

For that purpose, we firstly reduce the Eq.~\eqref{e:6} to a first-order Riccati differential equation, by substituting the field's time-dependence, $E(\tau)$
\begin{equation}\label{e:9}
E_1(\tau)=e^{\int(u(\tau)-f(\tau)/2)\,\d \tau}\,.
\end{equation}
The resulting Riccati equation for the function $u(\tau)$ is:
\begin{equation}\label{e:10}
\frac{\d u(\tau)}{\d\tau}+u^2(\tau)= -\frac{1}{2}\frac{\d f(\tau)}{\d\tau}+\biggl(\frac{f(\tau)}{2}\biggr)^2-1\,,
\end{equation}
assuming that $f(\tau)$ is known according to Eq.~\eqref{e:7}. There are no known solutions for such an equation in general.

We tested various representations of the functions $u(\tau)$ and $f(\tau)$, i.e. different representation of the electromagnetic field and the time-dependent conductivities, $E(\tau)$ and $\sigma(\tau)$. We found that if the function $f(\tau)$ is expressed through a function $b(\tau)$, such that
\begin{equation}\label{e:11}
f(\tau)=C\cdot  b(\tau)+\frac{1}{b(\tau)}\frac{\d\, b(\tau)}{\d \tau}-b(\tau)\cdot\int \frac{\d \tau}{b(\tau)}\,,
\end{equation}
($C$ is an arbitrary constant), an exact particular solution to the Riccati Eq.~\eqref{e:10} can be written as
\begin{equation}\label{e:12}
u(\tau)=\frac{C}{2}\cdot b(\tau)-\frac{1}{2b(\tau)}
\frac{\d\, b(\tau)}{\d \tau}-\frac{b(\tau)}{2}\cdot \int \frac{\d \tau}{b(\tau)}\,.
\end{equation}
The representation of the conductivity function $f(\tau)$ in terms of another function $b(\tau)$ in Eq.~\eqref{e:11} is just a substitution of one function with another, in order to get mathematical advantages. It is valid for any function and do not involve any assumptions on the functions. The resulting Riccati Eq.~\eqref{e:10} and Eq.~\eqref{e:12} are therefore exact equivalents of the wave equation Eq.~\eqref{e:6} for the electric component of electromagnetic field.

A particular solution for the electromagnetic field determined by Eq.~\eqref{e:9} is then
\begin{equation}\label{e:13}
E_\text{part}(\tau)=e^{\int(u(\tau)-f(\tau)/2)\d \tau}=e^{\int\left(-\frac{1}{b(\tau)}\frac{\d\, b(\tau)}{\d \tau}\right)\,\d \tau}=\frac{1}{b(\tau)}\,.
\end{equation}
To find a general solution for the electromagnetic field, the substitution $\bigl($in terms of a new introduced function $z(x)$$\bigr)$
\begin{equation}\label{e:14}
E_1(\tau)=E_\text{part}(\tau)\cdot z(\tau)=\frac{z(\tau)}{b(\tau)}
\end{equation}
transforms the initial differential Eq.~\eqref{e:6} for the field, with the conductivity-related function $f(\tau)$ expressed as in Eq.~\eqref{e:11}, into a first-order ordinary linear differential equation for the derivative of $z(\tau)$,
\begin{equation}\label{e:15}
\frac{\d^2 z(\tau)}{\d \tau^2}=\frac{\d z(\tau)}{\d\tau}\biggl\{-C\cdot b(\tau)+\frac{1}{b(\tau)}\frac{\d\, b(\tau)}{\d\tau}+b(\tau)\int \frac{\d \tau}{b(\tau)}\biggr\}\,,
\end{equation}
which can be solved easily. The solution of Eq.~\eqref{e:15},
\begin{equation*}
z(\tau)=C_2+C_1\cdot \int\limits_0^\tau \d \tau \,e^{\int\left(-C b(\tau)+\frac{1}{b(\tau)}\frac{\d\, b(\tau)}{\d\tau}+b(\tau)\int\frac{\d \tau}{b(\tau)}\right)}\,,
\end{equation*}
gives the following general solution for the time-dependent electromagnetic field:
\begin{equation}\label{e:16}
E_1(\tau)=\frac{C_2}{b(\tau)}+\frac{C_1}{b(\tau)}\cdot \int\limits_0^\tau\d \tau \,b(\tau)
e^{\int\left(-C b(\tau)+b(\tau)\int\frac{\d \tau}{b(\tau)}\right)}\,.
\end{equation}
The constants $C_1$ and $C_2$ are found from the initial conditions in Eq.~\eqref{e:8} to be:
\begin{equation}\label{e:17}
\begin{split}
C_1&=\biggl(i-C b(0)+b(0)\cdot \int\frac{\d \tau}{b(\tau)}\biggr|_{\tau=0}\biggr)\cdot
e^{-\left\{\int\d\tau\left(-C b(\tau)+b(\tau)\int\frac{\d \tau}{b(\tau)}\right)\right\}\bigr|_{\tau=0}}\,,\\
C_2&= b(0)\,.
\end{split}
\end{equation}

The solution for the field (Eq.~\eqref{e:16}) is an \textit{exact
  general solution} describing how the radiation field is transformed
by time-variations of conductivity as described in a general form by
Eq.~\eqref{e:11}. Advantages of exact analytical solution over
numerical solutions are much more profound for powerful fast processes
(i.e., mathematically, for large-amplitude rapidly varying
coefficients of differential equations at a time-scale comparable to
the wave period), as conventional numerical techniques fail to produce
correct reliable results in this case. We are interested in the case
when the conductivity's time-dependence has a flash-like shape,
describing localised ionisation or discharge events. We demonstrate in
Sect.~\ref{ssec:2.3} how we model the time-dependent conductivity by
the choice of a function $b(\tau)$ that describes a flash-like
ionisation event.

\subsection{Conductivity time-dependence for flash ionisation processes}\label{ssec:2.3}

The function $b(\tau)$ determines the time-dependence of the
conductivity by Eq.~\eqref{e:11}, as well as the solution for the
electromagnetic field by Eq.~\eqref{e:17}. We wish to describe the
conductivity flash variation most accurately by allowing it to be as
flexible as possible. We require our model to be able to describe two
different types of lightning discharges: a capacitor-like one and one
induced by runaway electrons. A time-dependence corresponding to a
runaways induced discharge, like that known for cosmic ray induced
lightning discharges~\citep{gurevich13}, starts as exponential
growth. Unlike a capacitor-like discharge, which would start with
almost linear growth and requires a high degree of charge separation,
runaway discharges require less initial potential difference (see
e.g.~\citep{hel13a}). The runaway-initiated discharges are therefore
more easy to start, subject to existence of a small population of
runaway electrons. The time-dependent conductivity of such initiated
discharges starts as an exponential growth of the conductivity due to
the breakdown of the avalanche.

\begin{figure}
	\centering
	\subfigure[]{\hspace{1mm}\includegraphics[width=0.98\linewidth]{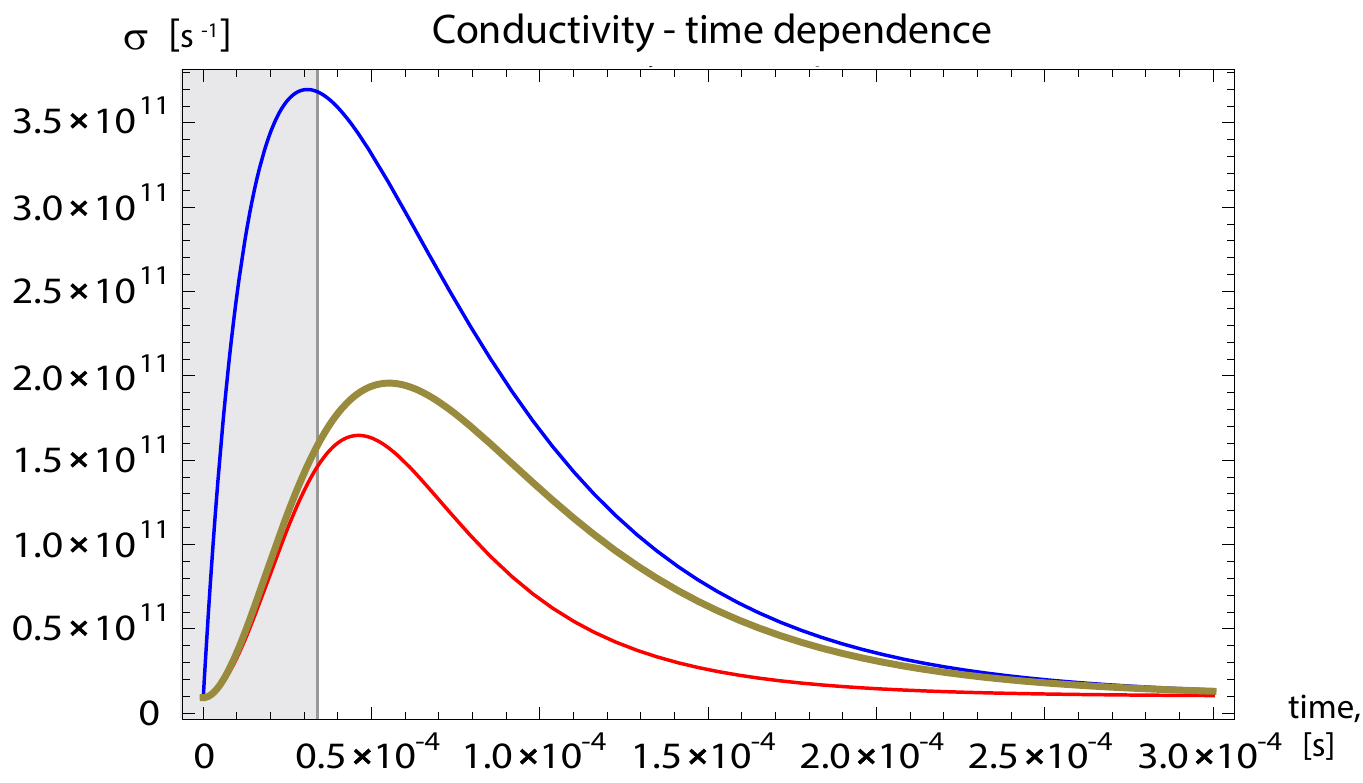}}
	\label{f:4a}
	\\
	\subfigure[]{\includegraphics[width=0.97\linewidth]{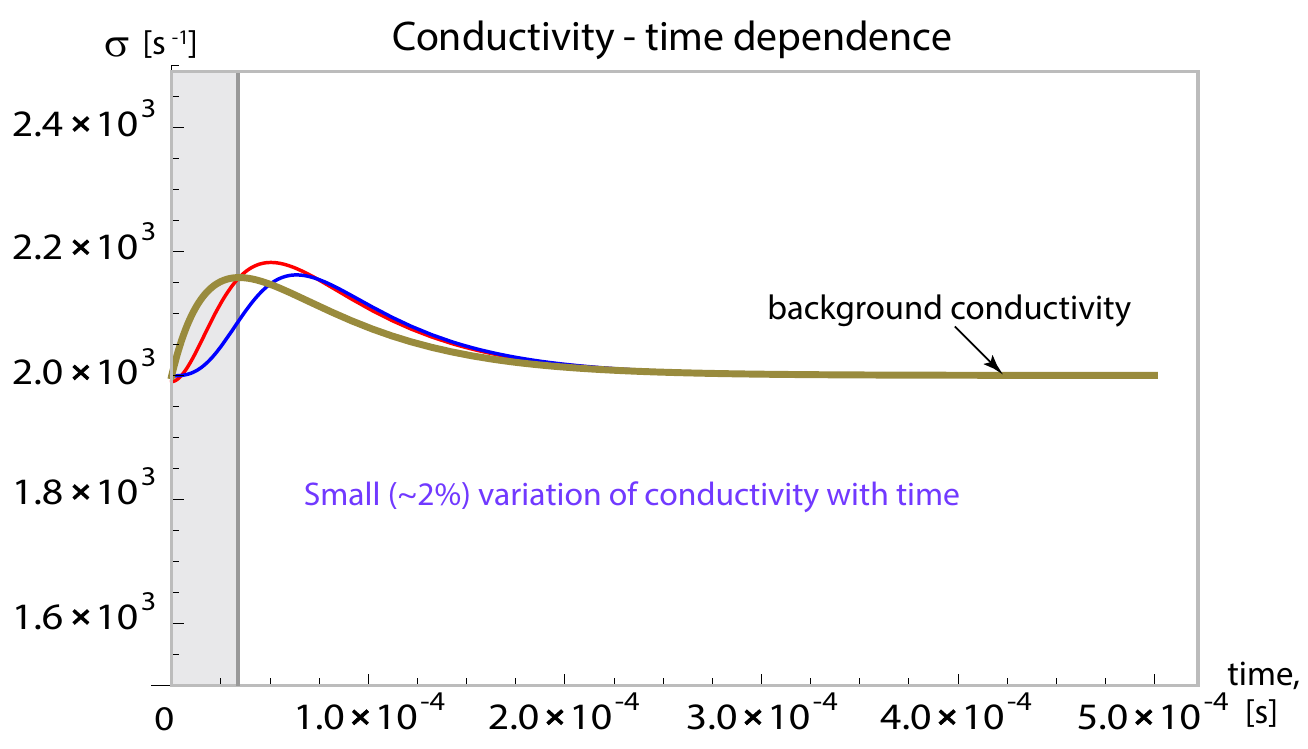}}
	\label{f:4b}
	\caption{Varieties of conductivity time-dependence shapes,
          described by the model in Eq.~\eqref{e:19}: (a) medium
          amplitude of the conductivity variation: capacitor-like
          discharge (blue line) and runaway-initiated breakdown,
          representing discharges triggered by cosmic rays (red and
          brown lines); (b) small variations of conductivity within
          10\% of its initial (background) value. Examples for
          parameters are given in Table~\ref{t:abcn}. Shaded regions correspond to the duration of the field response, which is a fraction of the conductivity flash's duration. }
	\label{f:4}
\end{figure}

We represent the time-dependent conductivity by the function $b(\tau)$ which allows us to describe both the capacity and the runaway breakdown discharges as part of the same formulae:
\begin{equation}\label{e:18}
b(\tau)=\frac{(\tau^k+B)^2 e^{b \tau}}{A\cdot \bigl(\tau^{2k}+(D+B)\tau^k+k(D-B)\tau^{k-1}+BD\bigr)}\,.
\end{equation}
The conductivity changing with time from Eq.~\eqref{e:10} is now
\begin{equation}\label{e:19}
f(\tau)=\frac{A(\tau^k+D)e^{-b \tau}}{(\tau^k+B)}+C\,.
\end{equation}
The parameters $A$, $B$, $C$, $D$, $b$ and $k$ are used to
describe the desired variety of the particular time-dependence's
shape. Figure~\ref{f:4} demonstrates various cases of the time
evolution of the flash-like conductivity that we are interested
in. The blue line in Fig.~\ref{f:4}(a) shows a capacitor-like
discharge, while two other curves demonstrate a runaway-initiated
discharge (like one triggered by cosmic rays). Figure~\ref{f:4}(a)
shows \textit{intermediate}-value conductivity flashes, while
Fig.~\ref{f:4}(b) shows small variations of conductivity within 10\% of
the background value, i.e. the value existing there before the flash
ionisation event (e.g., background thermal ionisation in Brown
Dwarfs). Equation~\eqref{e:19} allows for the analytical solution of
the problem of electromagnetic field transformation, while describing
the desired features of the conductivity variations. The general
solution for the electric field component is then described by
Eq.~\eqref{e:17} with the function $b(\tau)$ as given in
Eq.~\eqref{e:18}.

This general solution for the electric field component describes a
pre-existing radiation's response to a flash ionisation event. Such
events can take place in cloudy substellar atmospheres but also in,
for example, protoplanetary discs.

\subsection{Form of conductivity time-variations as a result of natural discharges}\label{ssec:3.4}

Time-depending conductivity correspondent to a flash ionisation
process could be expected in a form of a flash-like shape, firstly
rising due to fast growth of the ionised particles at a breakdown
ionisation stage, then reaching saturation and eventually declining
when the process is over. Direct and indirect (through measuring a
current) evidence of a flash-shape time-dependent conductivity can be
found in observations (e.g., on Earth ~\citet{lu10} or on Jupiter ~\citet{farrell99}) and experiments (e.g., nuclear lightning ~\citet{gardner84} or laboratory experiment, ~\citet{aleks9}). To check the
relevance of our conductivity-flash model to lightning discharges and
similar phenomena and to conclude on quantitative parameters which can
be expected for the conductivity flash, we look into the observational
results of a discharge associated with terrestrial gamma ray flash
(TGF) analysed in~\citep{lu11}. Detection of radio signals in ULF and
VLF was reported, with gamma rays detected within 0.2~ms of the fast
discharge. Similar distinctive variations of the waveforms were
reported for another TGF-associated intra-cloud
discharge~\citep{lu10}. TGF production was reported to be associated
with the upward propagating leader during the initial development of
compact (1.5--2~km channel) intra-cloud flash within 30~km of the
sub-satellite point.

We use the integral Ohm's law

\begin{equation}\label{e:30}
j(t)=\int\limits_0^t \sigma(t-t') E(t')\d t'
\end{equation}

to retrieve the conductivity variation in time, correspondent to currents and fields which resemble the observed waveforms~\citep{lu10,lu11}. $j(t)$ and $E(t)$ in Eq.~\eqref{e:30} are the electrons current and electric field, correspondingly. 

\begin{figure}
\includegraphics[width=0.95\linewidth]{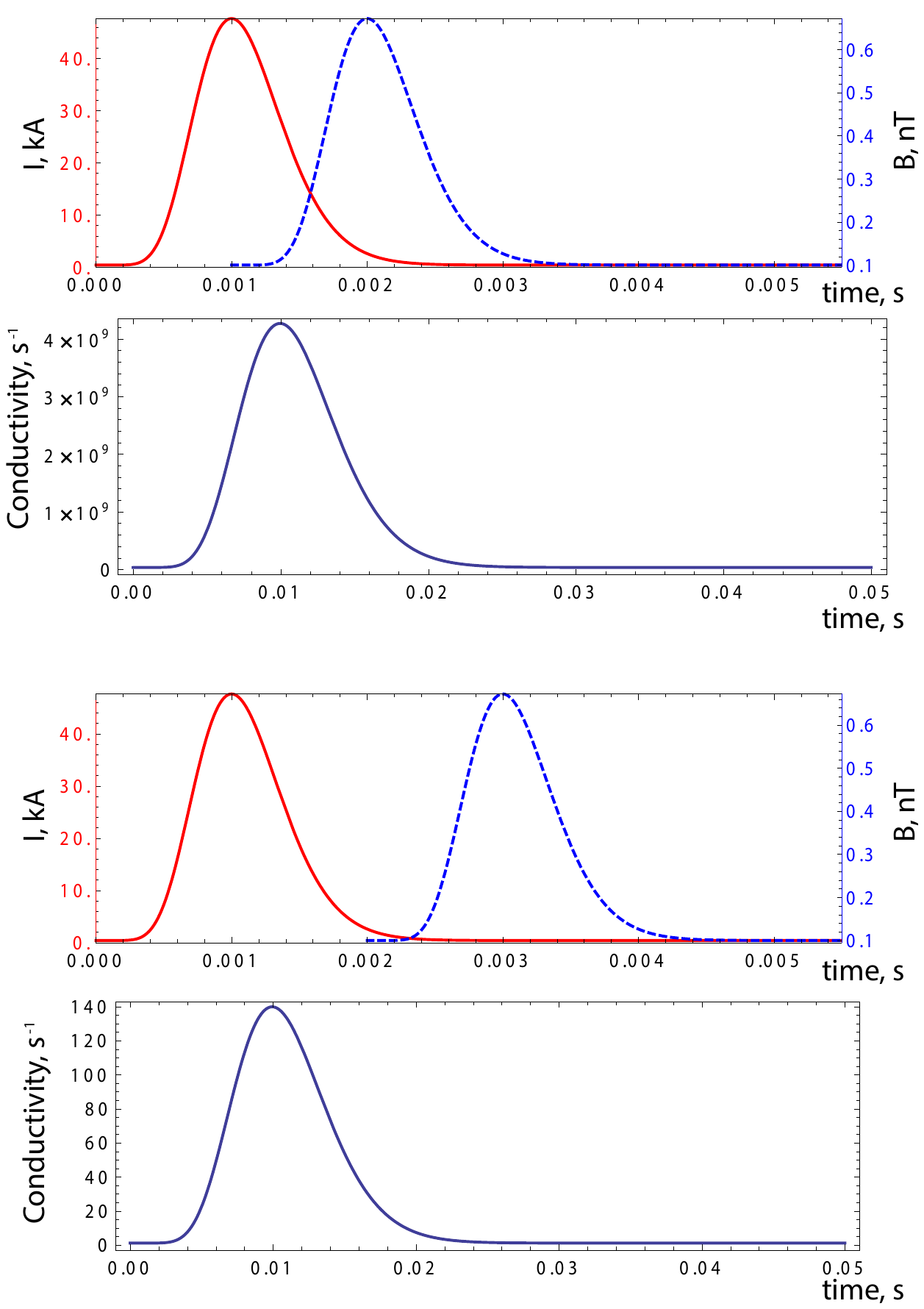}
	\caption{Waveforms of the electric current, {$I$\,[kA]} (red)
          and of the magnetic component of electromagnetic field,
          {$B$\,[nT]} (blue), with the retrieved conductivity time
          dependencies [1/s] for the correspondent $I$ and $B$
          combinations plotted below them. The electric current and
          the field have the waveform parameters similar to the
          observational results for a TGF-associated lightning
          discharge reported in~\citep{lu11}. Two different plots are
          for different delays of the electromagnetic production with
          respect to the current.}
	\label{f:9}
\end{figure}

The current and the field have a flash-like pulse waveforms in observations for the events like lightning discharges or TGFs. Often, this form is modelled with double-exponential representation \citep{farrell99}. We use the representation
\begin{equation}\label{e:31}
\left[
\begin{aligned}
&j(t)=j_0\cdot (t-t_0)^n \cdot e^{-\alpha t}+ c, \\
&E(t)=E_0\cdot (t-t_0)^m \cdot e^{-\beta t}+ C
\end{aligned}
\right.
\end{equation}
for them, which describes the pulse shape reasonably accurate and
allows analytical Laplace transforms, while not leading to divergence
in inverse Laplace transform, as a double-exponential representations
would do. The parameters $n$, $m$, $\alpha$ and $\beta$ as well as the
integration constants $c$ and $C$ are determined from reproducing TGF
associated lighting discharge from \cite{lu11} in Fig.~\ref{f:9}.

For known waveforms of the current and the field, Eq.~\eqref{e:30} becomes an integral equation for the conductivity $\sigma(t)$. The integral Ohm's law reflects the fact that the current and the field are not only affecting each other at any particular moment, but the full pre-history of their time-evolution has its influence.

Eq.~\eqref{e:30} is a Volterra type integral equation of a convolution type. This allows transforming it into an algebraic equation by Laplace transform, $t\to p$, and deriving then the conductivity as follows:
\begin{equation}\label{e:32}
\hat{\sigma}(p)=\frac{\hat{j}(p)}{\hat{E}(p)}.
\end{equation}

The time-dependent conductivity can be found from~\eqref{e:32} after applying an inverse Laplace transform,
\begin{equation}\label{e:33}
\sigma(t)=\hspace{-2mm}\int\limits_{-i\infty+\alpha}^{i\infty+\alpha} \hspace{-2mm} e^{pt}\frac{\hat{j}(p)}{\hat{E}(p)} \,\d p = \hspace{-2mm} \int\limits_{-i\infty+\alpha}^{i\infty+\alpha} \hspace{-2mm} e^{pt}\frac{\int\limits_0^\infty j(t')e^{-pt'} \d t'}{\int\limits_0^\infty E(t')e^{-pt'} \d t'}\,\d p.
\end{equation}

The current and the field in Eq.~\eqref{e:33} are determined by Eq.~\eqref{e:31}. 

We use the current and the field resembling those in \citep{lu10},
their Fig.3, to retrieve the conductivity waveform (its
time-dependence). Fig.~\ref{f:9} shows the current (red, solid) and
the field (blue, dashed) and the resulting retrieved conductivities
for two cases of different time delays between the field waveform
emission and the discharge current. While the shape of the
conductivity time variation does not depend on the time delay, its
peak value depends on it dramatically. Estimation of the time delay
however relay on accuracy of estimations for the distance from the
detectors to the discharge and for the correspondent velocities. The
duration of the retrieved conductivity flash does not depend on that
delay, and is almost ten times longer than that of the current or
field pulses (flashes). Similar results were obtained for the current
and field waveform reported in~\citep{farrell99} for Saturn
observations, but the conductivity flash duration is comparatively
shorter, being about five times longer than that of the current and
the field. Our model described by Eq.~\eqref{e:18},~\eqref{e:19} with
conductivity plotted in Fig.~\ref{f:4} is well representative of the
flash ionisation events like lightning discharges, as comparison with
the retrieved conductivity (Fig.~\ref{f:9}) suggests.

\section{The imprint of a time-dependent ionisation on a radiation field}\label{sec:3}

 Flash ionisation events like lightning, coronal discharges, energetic
 explosions or eruptions, can be suitably described by the
 time-dependent conductivity (Eq.~\eqref{e:19}), because this formula
 allows to adapt a suitable variation in amplitude, duration, shape
 and the flash front's gradient. We have demonstrated in
 Sec.~\ref{ssec:3.4} that our model does reproduce the well-studied
 case of TGFs' current (Fig.~\ref{f:4},~\ref{f:4}).

We consider the conductivity's temporal variations through flash
ionisation in a medium with a small value of natural background
conductivity, e.g. through thermal ionisation
(\citealt{Rodrigues2015}). This background ionisation is physically
required to start the flash ionisation. From a mathematical point of
view, the background ionisation is not necessary, and all the derived
formulae remain valid if the initial (background) conductivity is
zero. Figure~\ref{f:4}(a) shows larger amplitude changes in
conductivity, while Fig.~\ref{f:4}(b) demonstrates the changes within
a few percent of the background value.

The purpose of this paper is to study if and how the electric field
changes for different flash ionisation events (like coronal
discharges, lightning or sprites), which we represent by various
conductivity profiles (Eq.~\eqref{e:19}; Fig.~\ref{f:4}). As we aim to
find signatures of the flash ionisation events in pre-existing
cyclotron radiation, we consider radio frequencies as the incident
radiation field in our analysis, i.e. as a carrier signal for the
modulation imposed by a flash-ionisation. The analytical results from
the previous sections are valid for any frequency, i.e. for any
relation between the initial wave frequency and the time scale of the
ionisation process, as long as the assumption of a simultaneous
conductivity change in different points of the local medium is
reasonable.

\begin{figure}
\subfigure[]{\includegraphics[width=0.98\linewidth]{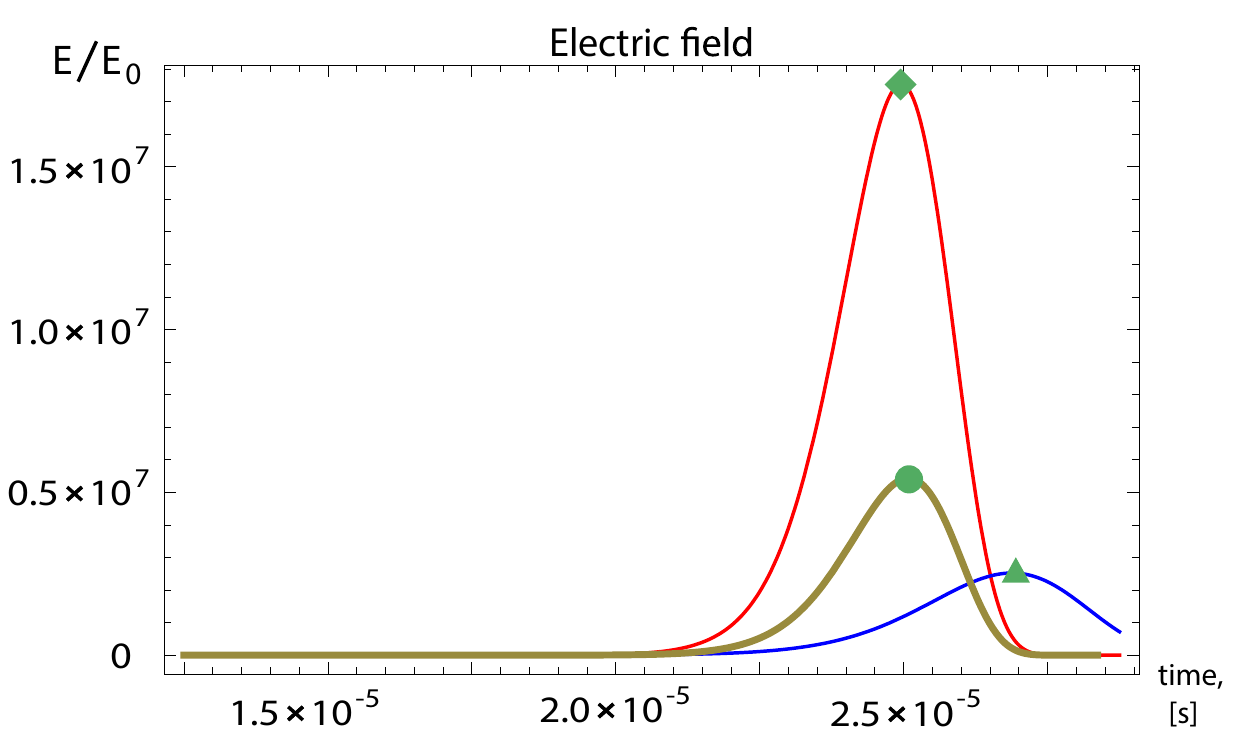}}
\label{f:5a}
\\
\subfigure[]{\hspace{5mm}\includegraphics[width=0.91\linewidth]{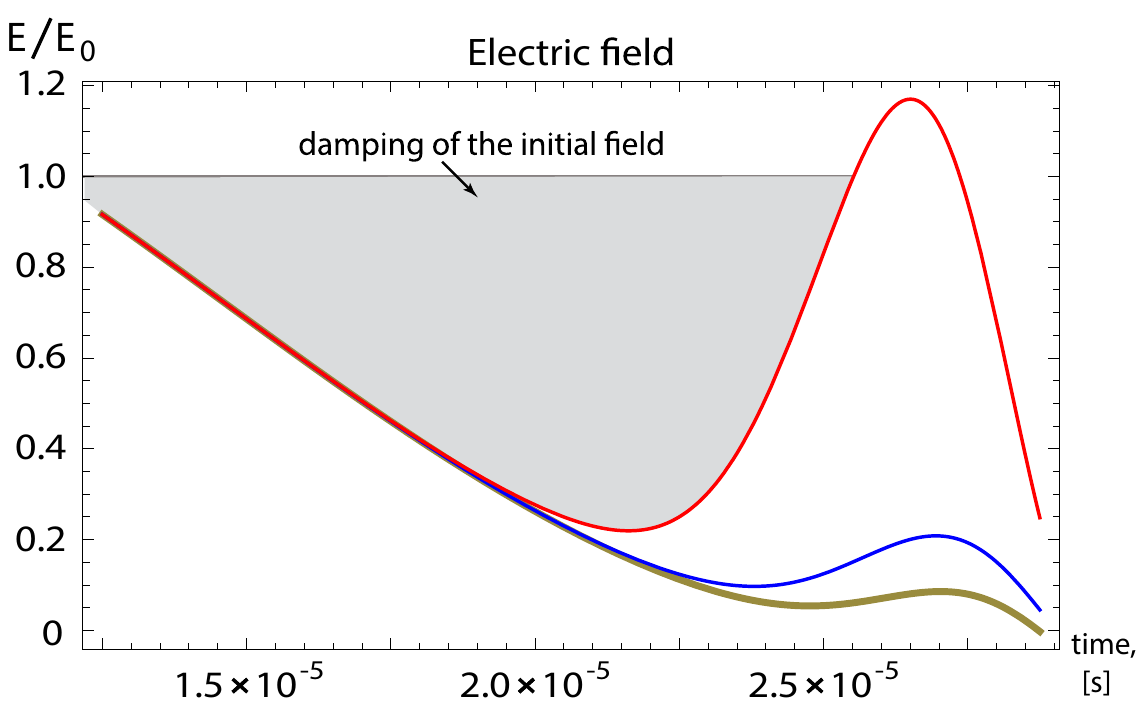}}
\label{f:5b}
\caption{ Typical electric field transformation (a) for
  \textit{intermediate} conductivity flashes; (b) for larger
  conductivity flashes. Colour code for the field is the same as for
  the conductivity time-dependences shown in Fig.~\ref{f:4}. Time
  interval on the plots corresponds to the grey area in
  Fig.~\ref{f:4}. The \textit{intermediate} conductivity flash results
  in a characteristic response in a form of a flash-shaped pulse of
  electric field, which can be substantially amplified with respect to
  the initial field amplitude~(a). The large amplitude conductivity flashes in
  time results in smaller field response, and for flash-character
  damping of the incident field~(b).}
\label{f:5}
\end{figure}

Using the model formulae presented in Sect.~\ref{sec:2},
Fig.~\ref{f:5} demonstrates how the electric field (Eq.~\eqref{e:17})
responds to a flash-like pulse due to a fast change of conductivity:
\begin{itemize}
\item  The response of the electric field begins when the conductivity
flash has a high positive gradient. 
\item The maximum of the field response
corresponds to the time of the maximum conductivity gradient.
\item  The
field responds on a significantly shorter time-scale than the
time-scale of the underlying process responsible for the rapid
conductivity change. 
\item The pulse duration of the response field is typically about ten
  times shorter than the conductivity flash duration.
\end{itemize}

For a fixed duration of the conductivity time-flash\footnote{the duration is taken at the half-height of the pulsed shape.}, the duration of the
response field pulse's vs the conductivity peak value is shown in
Fig.~\ref{f:7}. Remarkably,
\begin{itemize}
\item the
field response becomes shorter in time the more powerful the
ionisation process is (i.e. for a larger peak value of the flash's
conductivity).
\end{itemize}
 Figure~\ref{f:8} demonstrates that the field response
also starts slightly later for flashes with larger peak value of
conductivity, when making the comparison at the same fixed position of
the conductivity maximum. This can be attributed to the fact that the
maximum gradient of the flash is reached slightly later for larger
flashes. The duration of the responding electric field pulse
(Fig.~\ref{f:5}) is at least one order of amplitude shorter than the
duration of the ionisation flash.

\begin{figure}
\includegraphics[width=0.95\linewidth]{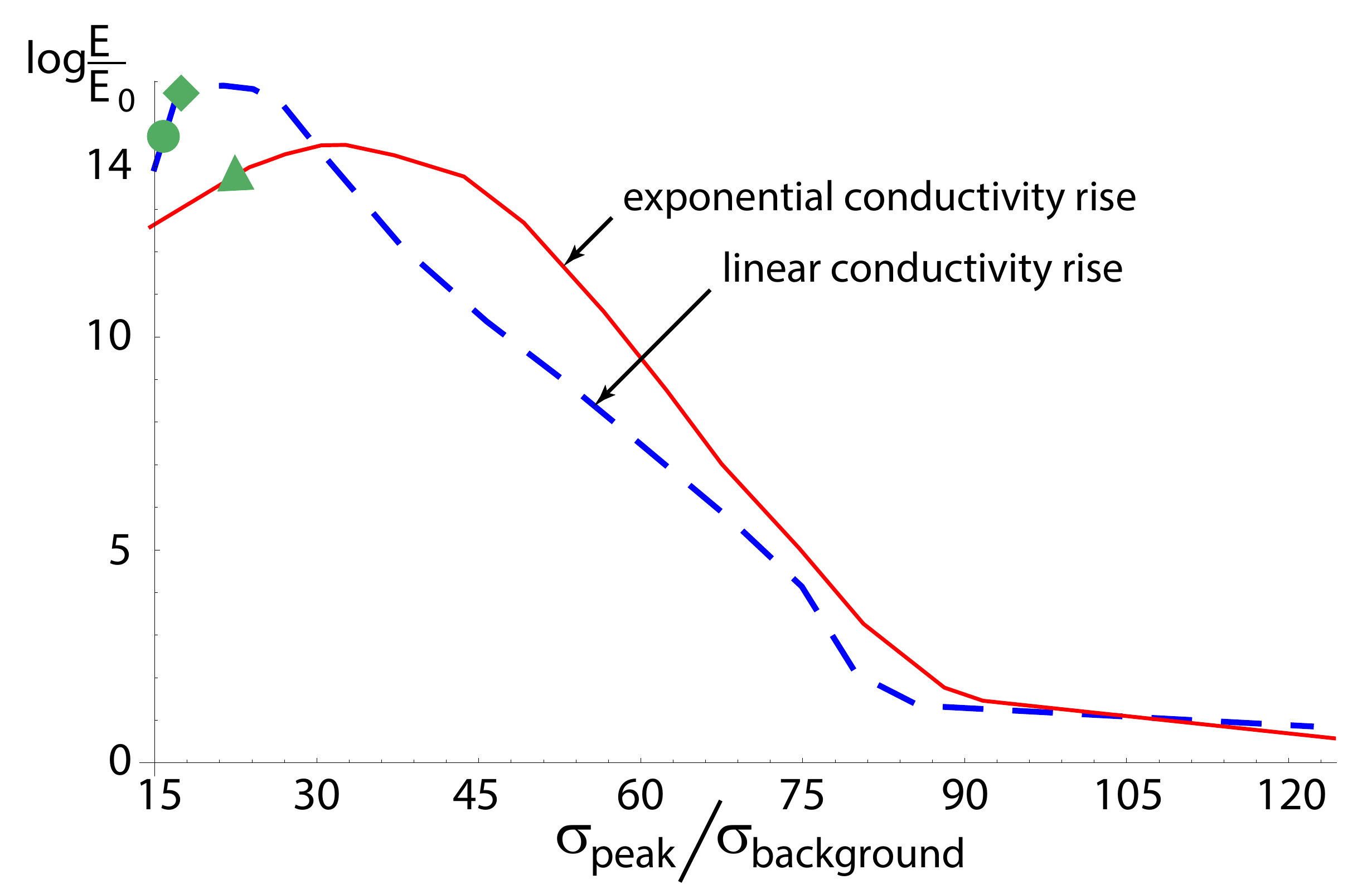}
\caption{The relative electric field amplitude, $\log(E/E_0)$, at its
  maximum value, vs the maximum (peak) value of relative conductivity
  change with respect to the background conductivity,
  $\sigma_\text{peak}/\sigma_\text{background}$. Dashed and solid
  curves correspond to cases of conductivity rising linearly up at the
  beginning of the flash process $\bigl($corresponding to the blue
  curve in Fig.~\ref{f:5}(a)$\bigr)$ and conductivity with exponential
  growth at the beginning of the flash $\bigl($corresponding to the
  red curve in Fig.~\ref{f:5}(a)$\bigr)$. One can see that bigger
  field response does not always correspond to bigger flash of
  conductivity (i.e. larger peak value of conductivity).}
\label{f:6}
\end{figure}

Investigating the dependence of the amplitude of the electric field
response pulse on the amplitude of the conductivity flash
(Fig.~\ref{f:6}) shows that for small variations of the conductivity
in time $\bigl($within a few percent, as shown in
Fig.~\ref{f:4}(b)$\bigr)$ the field's change is not noticeable.
\begin{itemize}
\item
 For \textit{intermediate} changes in time, however, the response amplitude
will be larger for more powerful ionisation (larger flash of
conductivity, see Fig.~\ref{f:8}).
\end{itemize}
 Note, that the conductivity can be described as sums of the products
 of densities for all available charges and the charges' mobilities,
\begin{equation}\label{e:21}
\mu=\frac{v_\text{drift}}{E_\text{electrostatic}}\,,
\end{equation}
where $v_\text{drift}$ is the electrons' drift velocity and
$E_\text{electrostatic}$ is the local electrostatic electric
field. During the avalanche (breakdown) process, the mobility of
electrons as well as their number density is much higher than those of
ions or charged dust grains, so we only take electrons motion into
account. The conductivity is then determined by
\begin{equation}\label{e:22}
  \sigma=e\cdot\mu_e\cdot n_e\,,
\end{equation}
where $e$ is a charge of the electron, $n_e$ is electrons' number
density and $\mu_\text{e}$ is the mobility of the electrons. A higher
peak value of conductivity requires therefore a combination of strong
ionisation and charge acceleration.

\begin{figure}
\centering\includegraphics[width=0.95\linewidth]{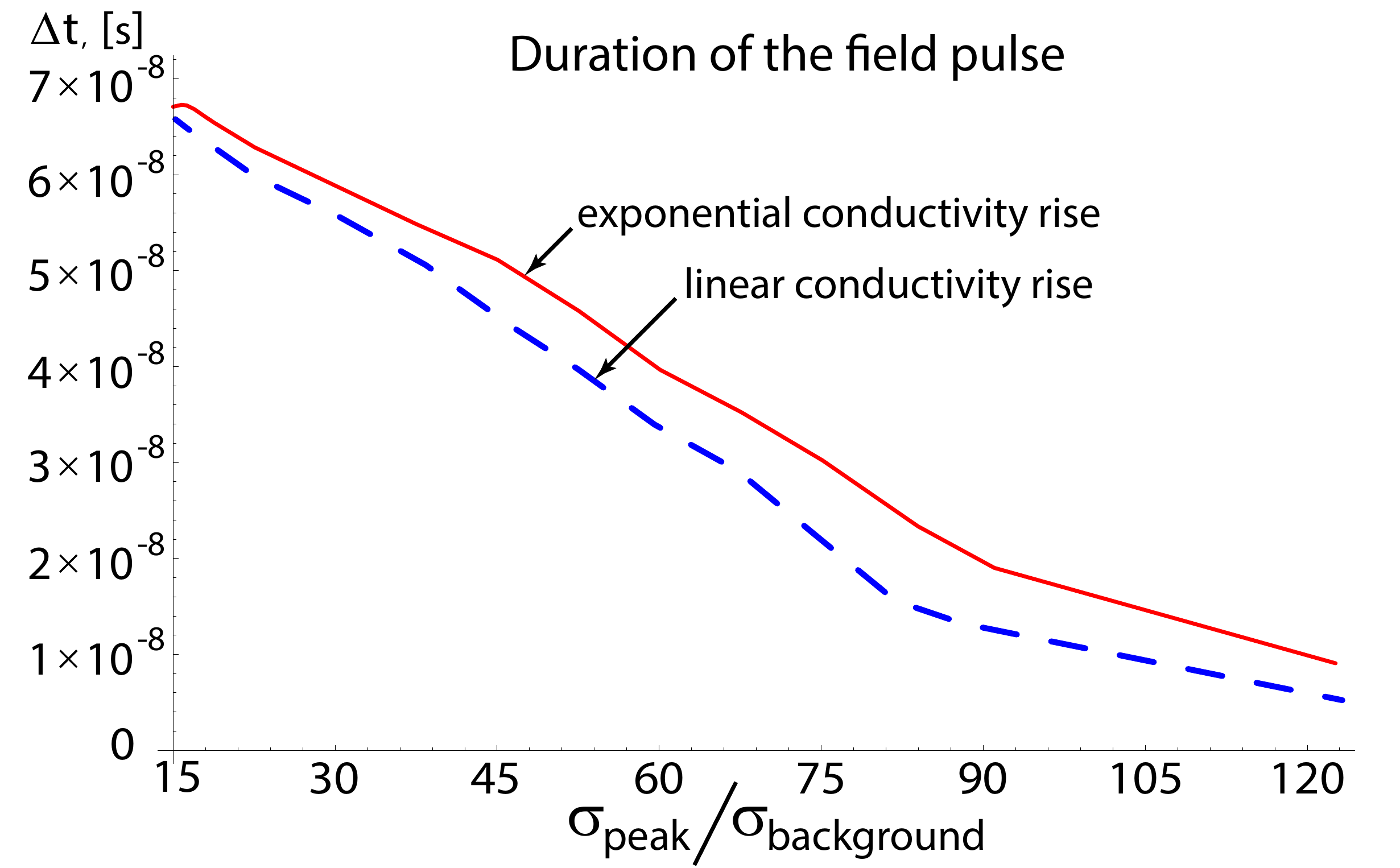}
\caption{Dependence of the electric field pulse duration on maximum
  value of relative conductivity change with respect to the background
  conductivity, $\sigma_\text{peak}/\sigma_\text{background}$. The
  duration is measured at $1/2$ of the field's maximum value at its
  peak. Colour code is the same as in Fig.~\ref{f:6}. The response
  field pulse is consistently shorter for larger flash of conductivity
  (larger peak conductivity value).}
\label{f:7}
\end{figure}

\begin{figure}
\centering\includegraphics[width=0.95\linewidth]{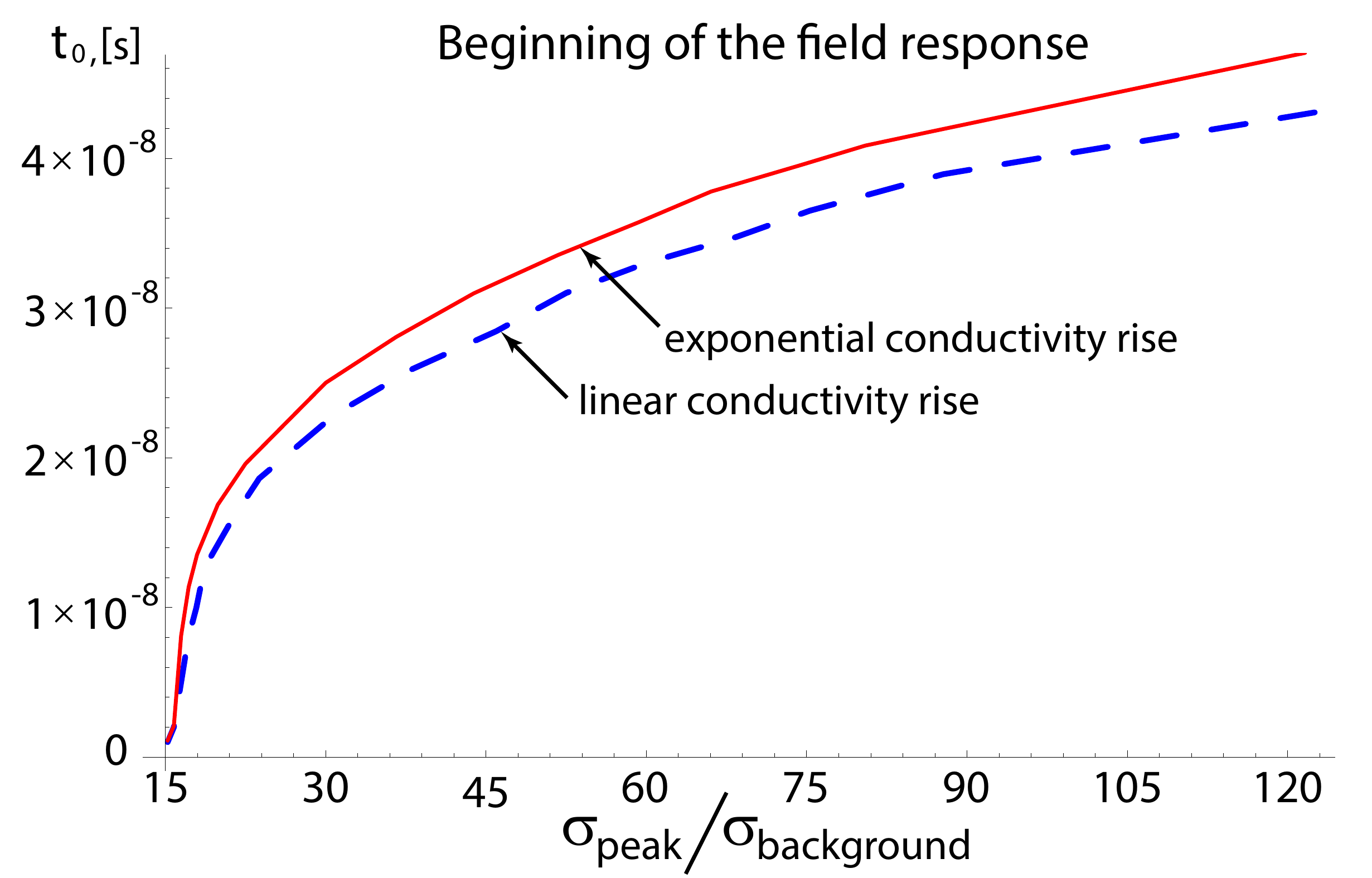}
\caption{Dependence of the electric field pulse start time, $t_0$, on
  maximum value of relative conductivity change with respect to the
  background conductivity,
  $\sigma_\text{peak}/\sigma_\text{background}$. The start time is
  determined on the field-rising slope at $1/2$ of the field's maximum
  value at its peak. Colour code is the same as in Fig.~\ref{f:6}. The
  field response starts consistently later for larger flash of
  conductivity (larger peak conductivity value).}
\label{f:8}
\end{figure}

For larger peak conductivities, the response field peak value becomes
less sensitive to the further increase of the conductivity, with the
dependence shown in Fig.~\ref{f:6} reaching saturation. Points shown
with a triangle, a circle and a diamond in Fig.~\ref{f:6} correspond
to the curves marked with the same symbols in Fig.~\ref{f:5}(a) and
lay within \textit{intermediate} flash ionisation
intensity. Increasing the peak conductivity even further (considering
more powerful flash ionisation processes) leads to reversal of the
effect: the field response becomes smaller for larger conductivity
flash, i.e. the normalised field amplitude, $E/E_0$ decreases with
increasing the relative conductivity change
($\sigma_\text{peak}/\sigma_\text{background}$).

It is known that lightning discharges radiate an electromagnetic field
(e.g.,~\citet{rycroft, rakov07, farrell99, uman64}). This radiation
occurs when runaway electrons, produced by ionising discharges, form
conducting currents which act as radiating antennae. The radiation is
produced during the energetic phase of the discharge, when the runaway
electrons are still fast enough to radiate~\footnote{mechanism of
  radiation here similar to radiation from an electric current.}, and
when their radiation is not yet fully damped by collisional
losses. Similarly, the short-term amplification of the pre-existing
radiation in a flash-ionised medium, as seen in Fig.~\ref{f:5}(a), is
caused by contributions of the energetic electrons' radiative
currents. Both, mobility and charges number density, influence
conductivity, as seen in Eq.~\eqref{e:22}. As noted in Sect. 1, the conductivity in our model\footnote{which is a
  general model for a transient electromagnetics as derived from
  Maxwell’s equations} is a combination of all effects from ionization
and charges movement, including the reactive and the dissipative
responses. The peak of the field amplification corresponds roughly to
the maximum gradient of the rising conductivity. This reflects on the
dynamics between two mechanisms of rising conductivity: the one
associated with increasing charges mobility and the one associated
with increasing number density of charges. Runaway breakdown at the
first stage of the discharge dramatically increases the charge
mobility aspect of the conductivity, while consequent continuing
increase in conductivity is caused mostly by the increased number of
charges from collisional ionisation. This second stage of the
conductivity rise will effectively damp the produced
radiation/amplification\footnote{Ionizing discharges' ability to
  produce a short-term field amplification is known and utilised in
  laboratory (e.g., \citet{bickerton58, skordoulis90}).}.

\textbf{}

While one might intuitively expect that a more powerful ionisation
process would lead to a higher response in the electromagnetic field's
amplitude, our results show that there is some threshold in the peak
value of conductivity (which is different for different initial
conditions and shapes of the conductivity time-flashes), corresponding
to the peaks, after which (i.e. for the conductivity change bigger
than that value) the field response decreases for larger flashes. The
reason for this is that, above a certain threshold, the conductive
medium would be efficiently damping the field, allowing the damping to
prevail over the effect of pumping the field by the transient event's
energy. An ionised medium (like the ionosphere) is generally
non-transparent to most of the radio frequencies. This is one of the
biggest problems with observing lightning discharges from outside of
the solar system or terrestrial planets. With additional energy coming
into the radiation field from powerful ionisation processes, such a
field has a chance to penetrate through even an initially ionised
medium. However, processes of higher extremity (involving larger peak
value of conductivity flash) would result in a metal-like ionised area
which would damp the electromagnetic field dramatically. The
classification is based on the effect which is produced by the
transient event on the electromagnetic field, e.g., `strong' event
(`big' conductivity flash) attributes to damping the field, as in
Fig.~\ref{f:5}b, while `intermediate' event denotes the flash
resulting in the field amplification, as in Fig.~\ref{f:5}a. Small
events (i.e., the ionisation events which change the conductivity
within a few percent of its initial value) do not produce noticeable
signatures.

The field's response amplitude cannot therefore be used as a single
criterion for the power of the underlying ionisation process, and most
powerful flash ionisation processes (including, for example,
discharges/lightning and explosions/eruptions) can be missed by
observations. The duration of the electric field response pulse,
however, provides the information about the magnitude of the
conductivity flash, as it universally decreases with the increasing
conductivity flash amplitude for the same fixed duration of the
conductivity flash.

\section{Observational effects of flash ionisation on radio signals from a Brown Dwarf}\label{sec:3.1}

\subsection{Frequency of the radiation}\label{sec:5.1}

We apply our model to the case of radio wavelength being the
pre-existing signal that can be modulated by a flash-ionisation
event. Our choice of parameters is guided by observations of cyclotron
radio emission from Brown Dwarfs and late M dwarfs. 4.8~GHz frequency
observations were reported from a late M dwarf~\citep{burgasser13},
and observations of radio emission from Brown Dwarfs in the frequency
range 4.4 -- 4.9 GHz were presented in~\citet{route}. \citet{hall07}
also observed cyclotron emission from Brown Dwarfs at 4.88~GHz and
8.44~GHz, and radio emission at 5.8~GHz was reported
by~\citet{will13}. The radiation frequency, being related to local
cyclotron frequency of electrons, is proportional to local magnetic
field. Brown Dwarfs' and other UCDs' magnetic field is in order of a
few kG~\citep{schrijver09}, providing the cyclotron radiation
frequency in order of a few GHz. A few Gauss planetary magnetic field
results in a cyclotron radio emission in tens kHz -- few MHz frequency
range (e.g.,~\citet{depater10} for Jupiter).

\begin{figure}
	\subfigure[]{\includegraphics[width=0.98\linewidth]{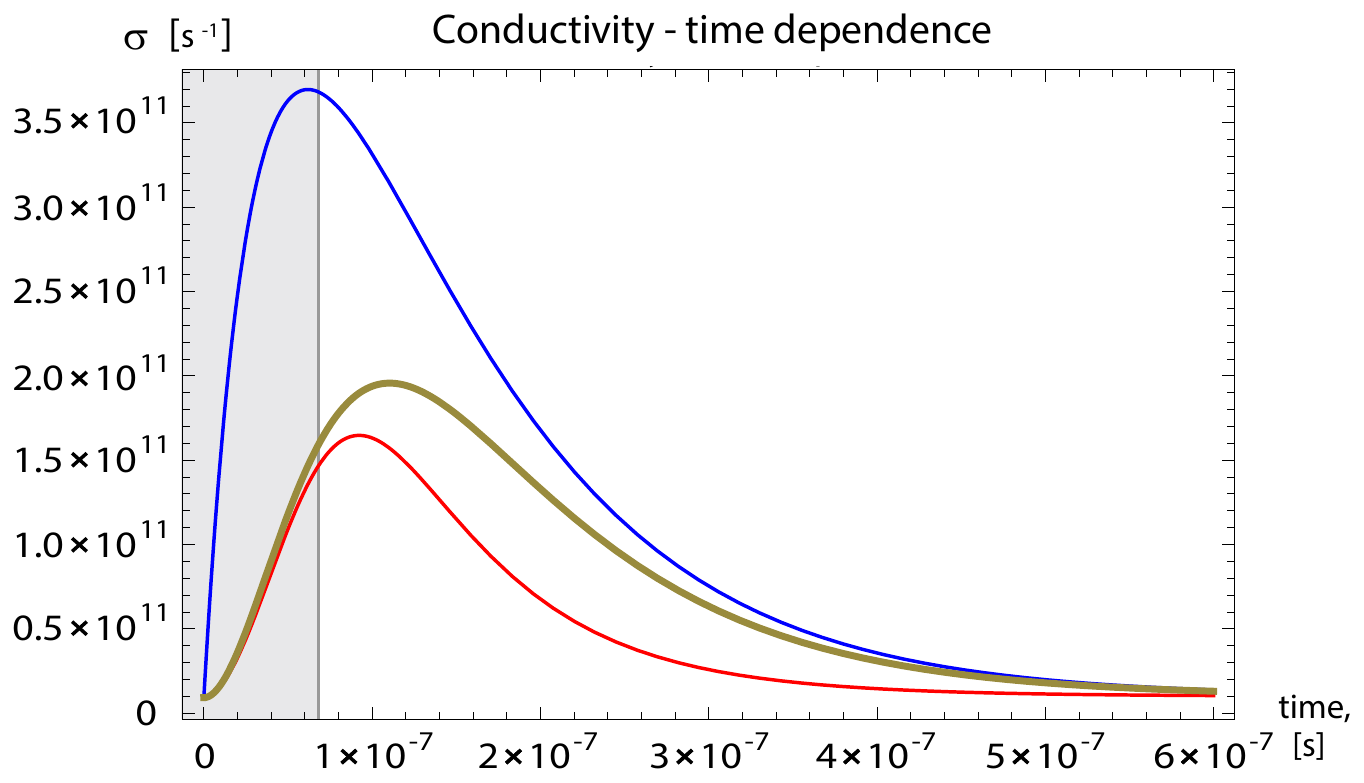}}
	\label{f:2a}
	\\
	\subfigure[]{\hspace{5mm}\includegraphics[width=0.91\linewidth]{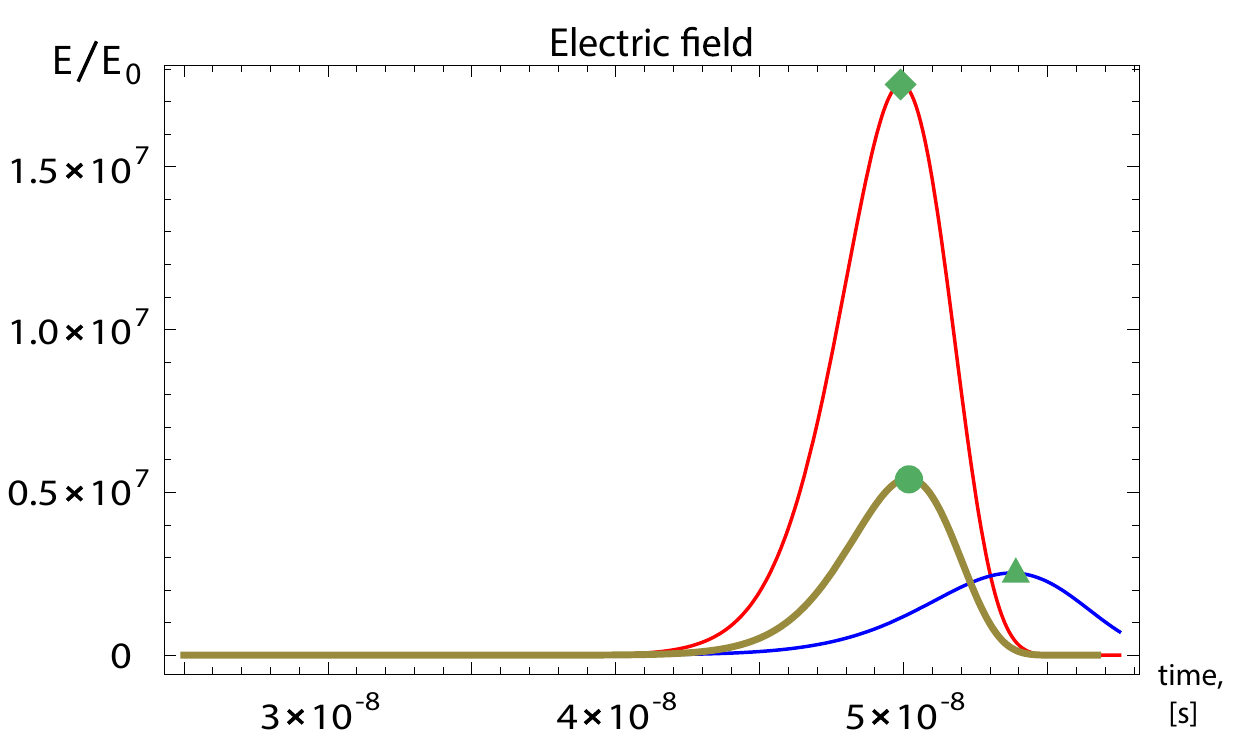}}
	\label{f:2b}
	\caption{
		(a) Conductivity time-dependence and (b) electric component of the electromagnetic field transformation for the frequency of initial radiation $f=1.59 MHz.$}
	\label{f:2}
\end{figure}

All our typical results are shown for $\omega$=4.8$\cdot 10^9~rad \cdot sec^{-1}$ (cyclic
frequency) in Figs.~\ref{f:5}-\ref{f:8}. They are representative for
other frequencies as well, as we demonstrate by Fig.~\ref{f:2} for the
alternative incident radiation frequency $f=\omega/2\pi = 1.59
MHz$(\footnote{Similar magnitude of the transformed field is achieved
  for about 500 times higher peak value of conductivity at the
  flash.}). The transformed electric field given by
Eqs.~\eqref{e:16}--\eqref{e:17} do not scale with the frequency
exactly. However, within the radio frequency range the temporal
characteristics of the field (start time and duration of the response,
Figs.~\ref{f:7},~\ref{f:8}) still roughly scale with the frequency as
$\omega^{-1}$, which makes our analysis more general.

\subsection{Peak value of conductivity}\label{sec:5.2}

Another key input value for the model of the flash is a peak value of
conductivity. We showed in Sect.\ref{ssec:3.4} that the peak
conductivity corresponding to terrestrial TGFs can be expected within
a wide range of values, depending on an actual delay time between the
measured current and field. For a cloud-to-ground discharge on Earth,
the peak value of temporal variation of the local temperature
$T_\text{gas}$ is typically about $T\sim
24{,}000$\,K~\citep{uman64}. A value for the conductivity derived by
these authors was calculated to be $\sigma \approx 1.08\cdot
10^{9}\text{s}^{-1}$.  \footnote{Found in \citet{uman64} conductivity
  is converted into CGS units as \begin{multline} \sigma\approx
    180\cdot 10^{-2} \ \Omega^{-1} \text{m}^{-1} \text{ (SI
      units)}=\\ 180\cdot 10^{-2}\cdot 9\cdot 10^9 \ \text{s}^{-1}=
    1.08\cdot 10^9 \ \text{s}^{-1}.
\end{multline}}
This is the conductivity that can be expected within an active
lightning channel on Earth. For the surrounding area, where the
lightning would produce flash ionisation of some (sometimes
substantial) volume, the conductivity is smaller than in the channel,
which could correspond to both `big' and `intermediate' flash regimes
(as defined in Sect.~\ref{sec:4}). However there are no values for the
peak conductivity for those areas in literature. Only the
`intermediate' flash effects can be seen by short-time amplification
as in Fig.~\ref{f:5}(a). However, the damping of the field as shown in
Fig.~\ref{f:5}(b) can also serve as an indication of an atmospheric
flash, though high intensity initial (cyclotron) radiation or high
signal resolution, sufficient for detecting 30\%--60\% variations in
amplitude $\bigl($such degree of damping of the initial field is
suggested by Fig.~\ref{f:5}(b)$\bigr)$ is required.

The conductivity of a medium can be calculated using
Eq.~\eqref{e:22}. Runaway electrons from the channel ionise the
surrounding medium. The conductivity is then proportional to the
electron drift velocity, $v_\text{drift}$. The velocity, and therefore
the mobility, is exponentially decaying with time~\citep{bruce41}
after the short initial period of electrons' acceleration by the
ionisation flash,
\begin{equation*}
\mu_e \sim v(t) \sim e^{-\gamma t}\,.
\end{equation*}
The number density of electrons is growing exponentially at runaway
breakdown before it reaches a saturation at about or after the moment
when the electrons start slowing down, $n_e \sim e^{\beta t}$. A
product of these two exponents, as in Eq.~\eqref{e:22}, one of which
is time-shifted (with the decay starting with some initial time delay)
and another is growing till the moment of saturation, results in the
flash time-dependence of conductivity. According to~\citep{rakov03},
the `burning point' at lightning discharge moves at $10^5$~km/s, while
the tip velocity of sprites from microdischarge model is
$10^4$~km/s~\citep{rycroft}. Exponential decay in the runaway
electrons velocity, \citep{bruce41} would result in an average peak
conductivity across the transient region of about $\sigma_\text{peak}
\sim 10^4$ s$^{-1}$. The $\sim 10^{-5}$~s duration of the conductivity
flashes was considered above as this would be within average values
for the duration of a cloud-to-ground discharge on Earth. The peak
conductivity of $10^4$ s$^{-1}$ corresponding to this duration is
small enough to be classified as an `intermediate' flash which
amplifies the field, as Fig.~\ref{f:5}a suggests.

With high-pressure atmospheres like those of BDs the number of
branches in lightning trees grow significantly with increasing
pressure \citep{bailey14, briels08}. This makes our model of a homogeneous
time-dependent conductivity more easily applicable to atmospheres of
Brown Dwarfs. Lightning on other planets and on Brown Dwarfs can be
more powerful and longer in time than on Earth, see

e.g.~\citep{farrell}. The duration of the flash ionisation of the area
surrounding the lightning channel can also be expected to be longer
than the lightning flash itself. For longer conductivity flash, the
peak conductivity as $\sigma_\text{peak} \sim 10^4$ s$^{-1}$ would
still be within `intermediate' flash limits, being therefore
detectable by the field amplification. Longer duration of the flash
ionisation process would also mean better chance of detecting the
response signal under the same time-resolution of detection. Thus,
high time-resolution of detection would be sufficient for detecting
highly amplified pulse response of the field from `intermediate'
conductivity time-flash, while both high time resolution as well as
intensity resolution (and so, high signal-to-noise ratio) are needed
to detect stronger flash events.

Powerful flash ionisation events like TGFs, are expected to produce
effects of `big' flashes, according to the high peak conductivity
values as in Fig.~\ref{f:9}. They are therefore detectable only by the
field's damping.  Note, that when mentioning `big' and `intermediate'
flashes of conductivity, we compare the conductivity peak values for
the same duration of the conductivity flashes. For the duration being
in order of $10^{-5}$ s the boundary between the `strong' and the
`intermediate' flash events is around the conductivity peak values at
few thousand $s^{-1}$. For longer duration, higher conductivity peak
values are required to classify the flash event as 'strong'. 

\begin{figure*}
\subfigure[]
{	
\includegraphics[trim=0.0cm 5.5cm 0.0cm 2cm, width=0.8\linewidth]{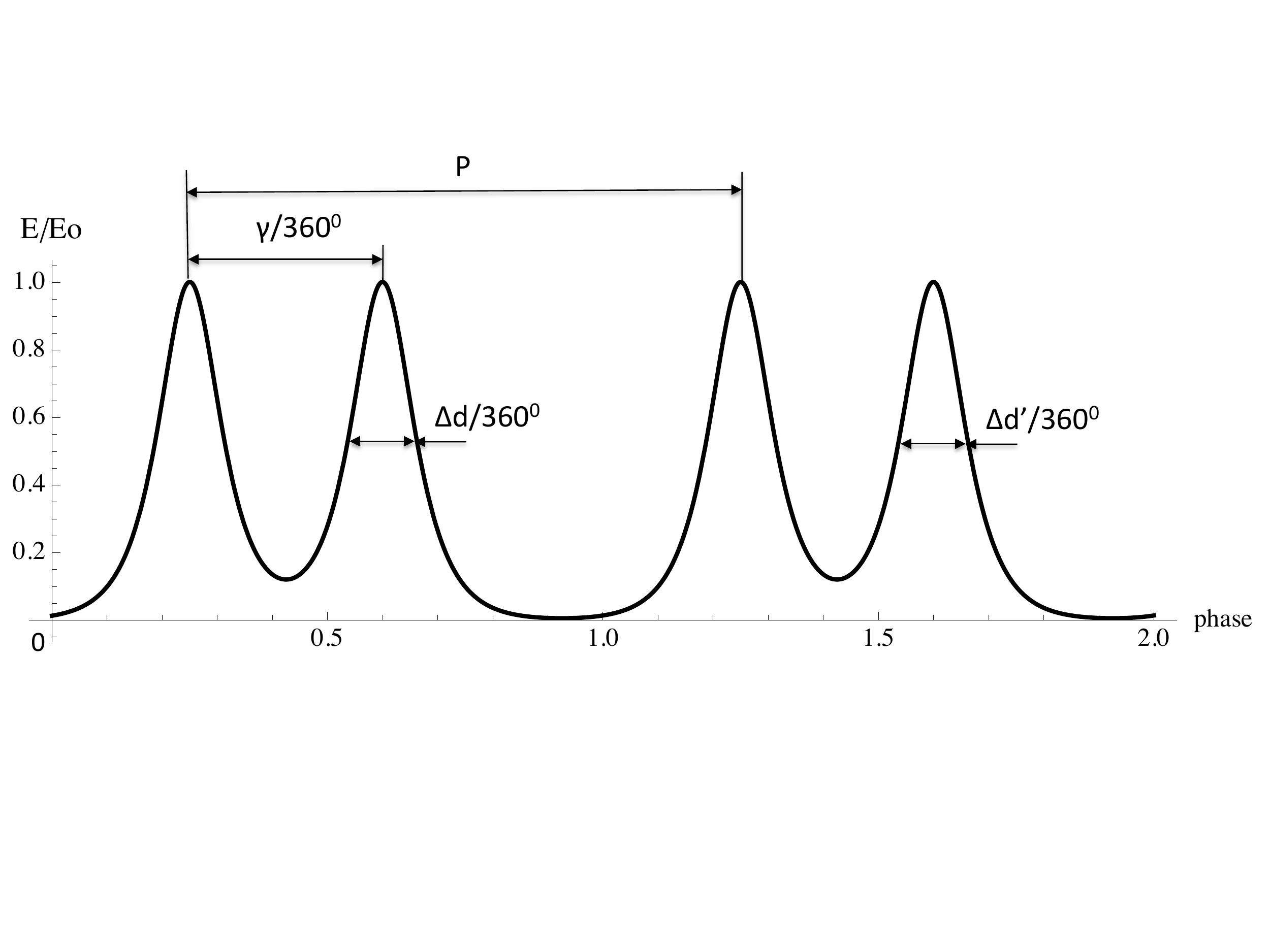}
}
\\*[-0.0cm]

\subfigure[]
{
\includegraphics[trim=0.0cm 0.0cm 0.0cm 1.5cm, width=0.5\linewidth]{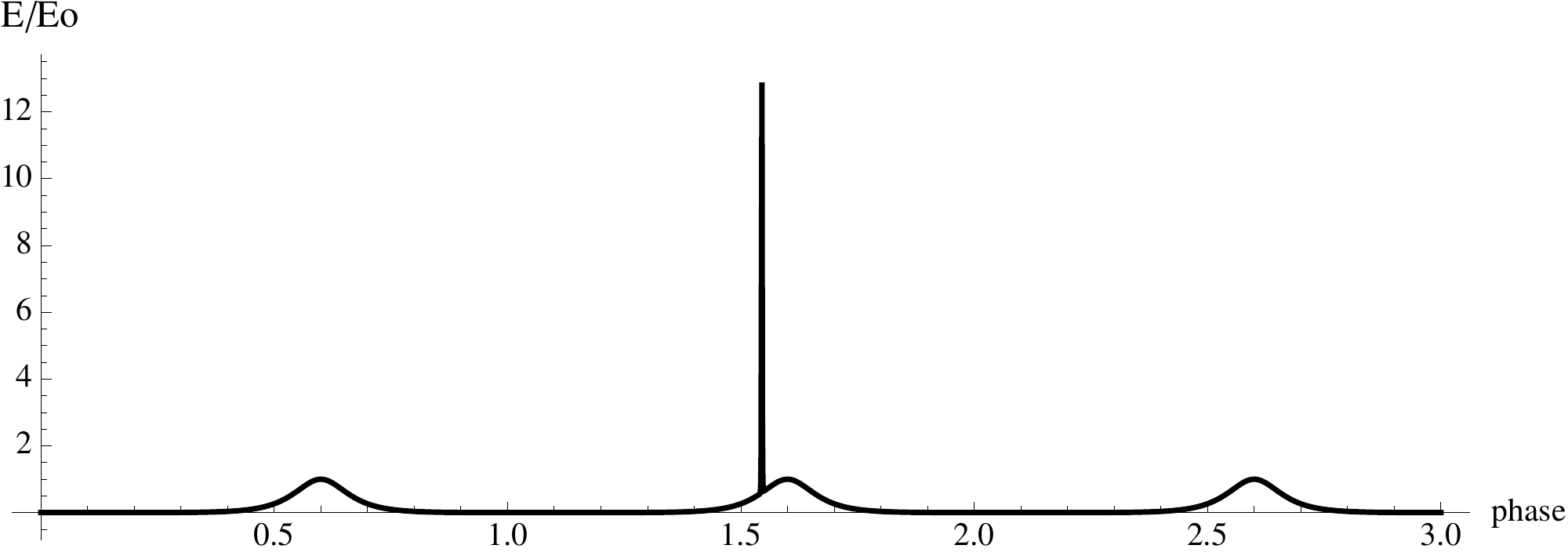}
}
\\
\subfigure[]
{
\includegraphics[trim=0.0cm 0.0cm 0.0cm 1.5cm, width=0.5\linewidth]{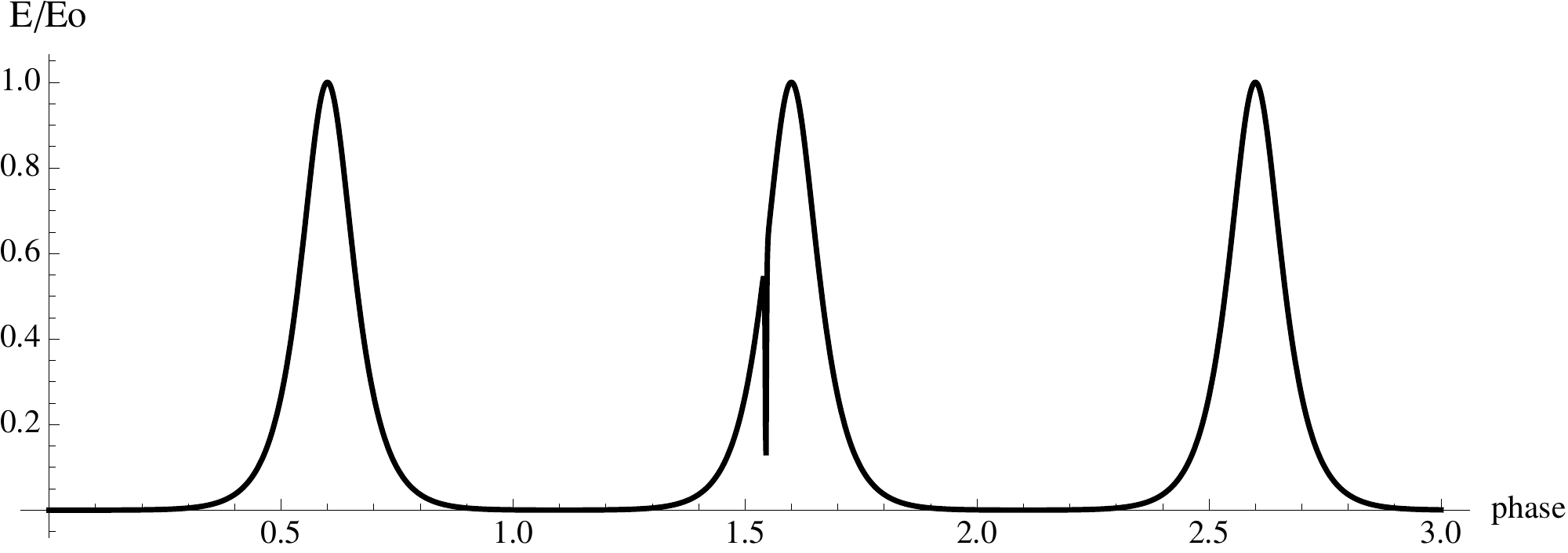}
}
\\
\subfigure[]
{
\includegraphics[trim=0.0cm 0.0cm 0.0cm 1.5cm, width=0.5\linewidth]{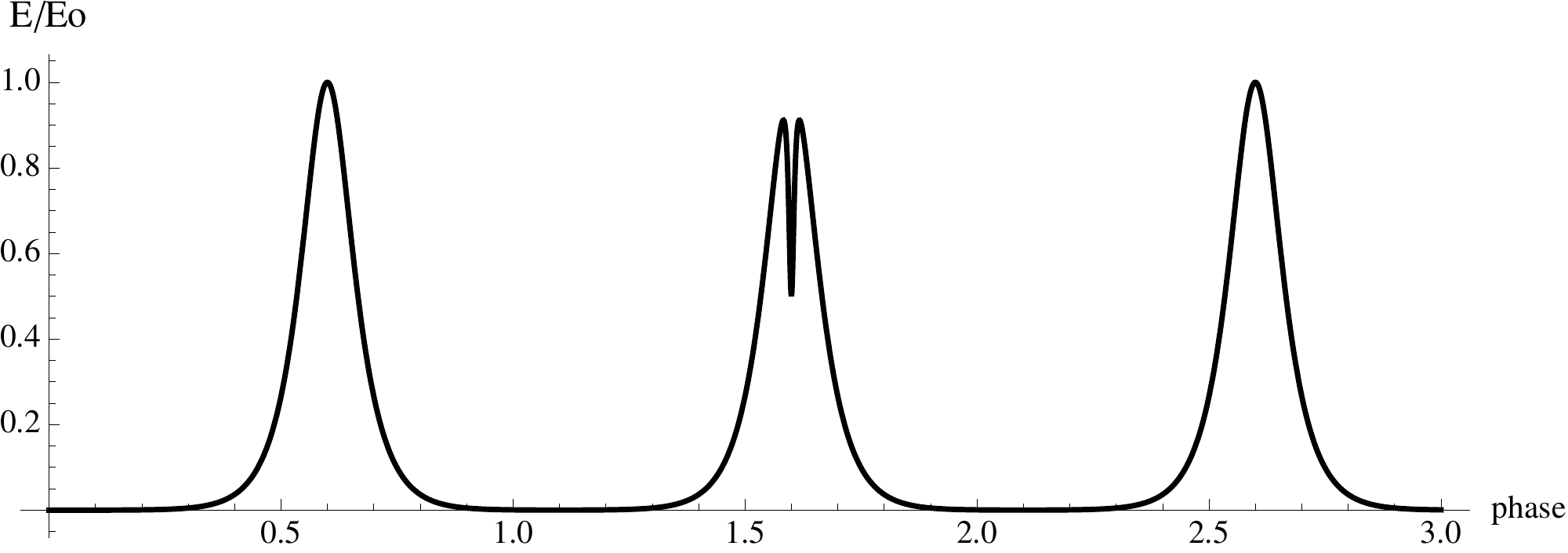}
}
\\
\subfigure[]
{
\includegraphics[trim=0.0cm 0.0cm 0.0cm 1.5cm, width=0.5\linewidth]{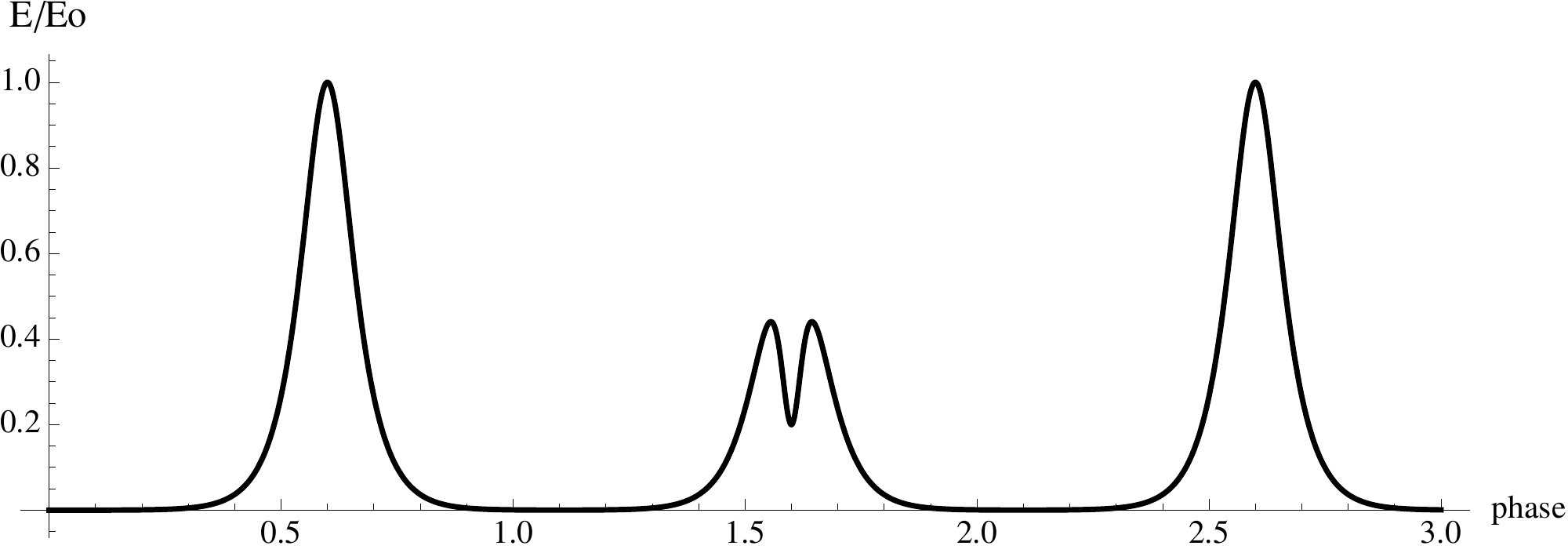}
}
	\caption{Time evolution of the electric field magnitude,
          $E(t)$, in units of the magnitude of the unperturbed field,
          $E_0$.  {\bf Top:} Full two-peaked radio signal from the two
          cone-walls with an observed time delay, $\gamma/360^o$
          related to the cone opening angle, and a beam width, $\Delta
          d$, which is equivalent to the emission cone wall thickness
          (compare Fig.~\ref{f:11}). The double-peaked signal repeats
          after a rotational period, $P$. The unperturbed case is
          shown where $\Delta d=\Delta d^{\prime}$. Time-dependent
          changes in the atmospheres and flash-ionisations can cause
          $\Delta d\not=\Delta d^{\prime}$ {\bf Bottom:} Examples of a
          periodically seen cyclotron emission (top) transformed by a
          flash ionization event as function phase. Only the changing
          amplitudes for one emission beam are shown.\newline  {\bf Bottom 1st row:}
          ({\it case (a)}) Amplification of electric field amplitude
          through intermediate conductivity flash peak values as in
          Fig.~\ref{f:5} (top).  {\bf Bottom 2nd-4th row:} ({\it case b - d}): Damping of
          electric field through large conductivity flashe peak values
          as in Fig.~\ref{f:5} (bottom).}
	\label{f:1}
\end{figure*}

\begin{table*}
\centering
\begin{tabular}{lllrr}
\hline
cases in Fig~\ref{f:1}: & a & b & c& d\\
\hline
\multicolumn{5}{l}{Flash parameters (Eq.~\ref{e:19}):}\\ 
A & $2.50\cdot10^{12}$ &  $2.50\cdot10^{13}$ &  $3.50\cdot10^{13}$ &  $4.00\cdot10^{13}$\\ 
B & $1.44\cdot10^{2}$  & $1.44\cdot10^{2}$   &  $2.304\cdot10^{3}$ &  $2.304\cdot10^{3}$\\
C & $3.00\cdot10^{9}$  & $3.00\cdot10^{10}$  & $4.00\cdot10^{10}$  & $4.50\cdot10^{10}$\\
D & -1.44& -1.44 & -23.04 & -23.04\\
$b$ & $8.33\cdot10^{-1}$& $8.33\cdot10^{-1}$ & $2.083\cdot10^{-1}$ & $2.083\cdot10^{-1}$\\
k & 2.0 & 2.0 & 2.0 & 2.0\\
\hline
\multicolumn{5}{l}{Flash properties:}\\ 
duration $\Delta \sigma$ [s] & 30 & 30& 120 & 120\\
peak value $\sigma_{\rm max}$ [s$^{-1}$]& $5\cdot10^{11}$  & $5\cdot 10^{12}$ & $7\cdot10^{12}$ & $8\cdot10^{12}$\\
 & intermediat & large & large & large\\ 
 & flash       & flash & flash & flash\\

\hline
\multicolumn{5}{l}{Electric field response (Fig.~\ref{f:1}):}\\ 
           & amplified & damped & damped & damped \\
\hline
\end{tabular}
\caption{Flash parameters for example cases of cyclotron emission
  changes due to flash ionisation encounter. These numbers are to
  visualise the effect and will change for more realisting beam
  functions, mulit-dimensional scattering effects etc. The flash
  duration $\Delta \sigma$ is taken at FWHH. }
\label{t:abcn}
\end{table*}

\subsection{Temporal variations of the signal}\label{sec:5.3}

The discussed above signatures of the cyclotron emission
transformation by a flash ionization event can be expected in a form
shown in Fig. 3 in the case when the emission is observed
continuously. Due to high directivity of the cyclotron emission and to
the emitting body’s rotation, this can only happen if the geometry of
rotation allows the radiated beam/cone staying at the same place with
respect to the observer, or if the observing station is following the
body’s (planet’s) rotation.

 In most of the cases the emission is seen as periodic bursts at the
 phases when the beam turns towards Earth
 (e.g.,~\citet{hall07,doyle10}), as in Fig.~\ref{f:1}, top. If a flash
 ionization event happens on the path of one of the periodic beams
 (Fig.~\ref{f:11}) out of the atmosphere, it would not change the
 periodicity but influence the signal's amplitude. 

 Figure~\ref{f:1} visualises four example cases how flash ionisation
 can alter a cyclotron signal. The beam shapes were calculated from
 $E(t)/E_0=\mbox{sech}[10\cdot(2t-1)]+\mbox{sech}[10\cdot(2t-3)]+\mbox{sech}[10\cdot(2t-5)]$
 ($\mbox{sech}(t)=(cosh(t))^{-1}$, t -- time in units of phase) but
 any suitable function (like Gaussian) could be used. Please note that the plots in Fig. 9 do not include rigorous modelling of the propagation of the transformed (by flash ionization) signal towards the observer. Instead, a model modulation function (imitating effects of the emission source’s rotation) shown in Fig. 9 (a) is used. The plots in Fig. 9 (b-d) are obtained as products of the field transformed by the transient event and the modulation function. This way, the plots represent predictions for the observations, while plots in Fig. 4 present the transformed field at the source.The top panel in Fig. 9
 summarizes the ideal, unperturbed situation where two emission beams
 reoccur after one rotational period, $P$. The half-height width of the peaks is
 determined by the wall thickness, $\Delta d$, of the emission cone as
 shown in Fig.~\ref{f:11}. The time interval between the two beams,
 i.e. between the emission maxima, is related to the emission cone
 opening angle, $\gamma$ (Fig.~\ref{f:11}). The changing peak height
 and width of the cyclotron emission due to an encounter with a
 lighting flash, for example, is show in the second and third row in
 Fig.~\ref{f:1}. These figures result from solving
 Eqs.~\ref{e:16},~\ref{e:17} for the electric field component applying
 a parametrised conductivity function (Eq.~\ref{e:19}, see also
 Sect.~\ref{ssec:2.3}). The necessary parameters for the presented
 results are summarized in Table~\ref{t:abcn}.

 Depending on the peak value of the conductivity flash induced by the
 ionisation and on its duration, the field modulation by the flash is
 either a short powerful burst of radiation (sharp increase of the
 amplitude, as in Fig.~\ref{f:1} ({\it case (a)}), or a damping of the whole
 radiation peak , Fig.~\ref{f:6} ({\it cases (c), (d)}), or its fragment,
 Fig.~\ref{f:1} ({\it case (b)}). The transformation of
 cyclotron emission which is seen in observations of JO746+20
 in~\citet{route}, their Fig. 5.01, looks similar to shown in
 Fig.~\ref{f:1} and can potentially be a sign of flash ionisation
 events affecting the cyclotron emission.

Detecting a signal pattern as shown in Fig.~\ref{f:1} can be
indicative of flash ionization events in the object’s atmosphere and
magnetosphere (like lightning or TGF). The cyclotron emission is
produced by an electron beam moving along the star/dwarf/planet’s
magnetic field lines. Its power depends on the beam’s properties,
including its geometrical properties, and number density and velocity
distributions. Significant amplification of a regularly seen emission,
like shown in Fig.~\ref{f:1}~(b) and ~\ref{f:1}~(c), is unlikely to
happen due to sudden changes in the beam, because of unrealistically
high amplitude for cyclotron emission. Damping of the whole signal or
its fragment, however, could also be caused by the electron beam
variations. Simultaneous optical observations can help to exclude or
confirm the beam’s variations, as visible aurora is induced by the
same beam as the radio cyclotron emission. Variation seen in the
visual aurora's geometrical pattern and intensity are expected to
correlate with variations in radio emission from the same beam. The
radio waves refracts differently compared to optical waves, hence,
radio ways will travel through different part of the atmosphere than
optical waves. Therefore, if the radio waves pass through a localised
flash ionisation event, the optical light may miss it. In a situation
when the visible light is still affected by the same event, this
effect should be smaller than that for radio emission, because the
visible light is emitted and propagates in a thicker cone. Therefore
the observed visible signal is an integral effect of the light
emerging from a significantly larger external surface of the
atmosphere than that of a directed radio cyclotron beam. This way,
only a small portion of this visible light can be affected by a local
event, making the effect of flash ionisation on visible light smaller
compared to its effect on radio waves.

Another reason for cyclotron emission variability is spin-modulation
of the emission due to a character of the star/planet's rotation. In
particular, when a planet's magnetic dipole is offset from its
rotation axis, the solar/star wind electrons at different rotation
phase have different paths towards the magnetic poles. This results in
the emission produced at different altitudes, where different values
of magnetic field account for the emission
variability~\citep{morioka}.

\subsection{Time scale of the field response}\label{sec:5.1}

The duration of the response field pulse (including both cases of
amplification and damping in the pre-existing radiation waveform) is
determined by duration,$\Delta \sigma$, of the conductivity flash. As
discussed in Sect.~\ref{ssec:3.4}, for the single discharges in Earth
atmosphere the response field pulse is about ten times shorter than
the duration of the discharge. For Saturn it was found that it is only
five times shorter \citep{farrell99}. Though single branches of lightning in Earth atmosphere are known to last for a few $\mu$s to a few $ms$, a lightning event consists of a tree of strokes, coming as groups of branches. \citep{pawar04}, studying multiple-discharge flashes in the atmosphere of Earth, found that  flashes in each group are bunched together for ~15–-20 min. There are also reasons to expect longer discharges in extraterrestrial atmospheres with different compositions and higher local pressure. The duration of discharges in air is limited by electron attachment because free electrons are easily lost in air through attachment to oxygen. Duration of discharges in other gases and under higher than Earth atmospheric pressure can be significantly longer than in air, as proved by experiments in \citep{briels08}. The field’s response stretches in time proportionally to the longer integral flash duration, reaching about 5-10 folds of the flash’s duration, which makes it observable with current instrumentation for large/average expected flashes of lightning. A time resolution on a scale shorter than the response
pulse duration is needed to follow the shape of the response filed
pulse (waveform). However, for its detection, even one measurement
within its duration will be indicative (Fig.~\ref{f:1}, lower panels)
which can be obtained even when the resolution is lower than the
response duration. This makes the detectability subject to the
statistical occurrence of transient ionisation events which is well
studied on Earth, and to a lesser extend for the solar system
planets. From Fig.~\ref{f:1} one also concludes that for an
  example period of $P\approx4$ hours a large flash would cause a
  substantial damping of the emission peak by $\approx 50\%$ with a
  time-dip of $\approx 3.2 ~min$ ({\it case (d)} in Fig.~\ref{f:1}). Intermediate
  flashes do produce a substantial amplification over a very short
  time window ({\it case (a)} in Fig.~\ref{f:1}).

\subsection{Recipes for observations}\label{ssec:5.1}

We have shown that transient ionisation events can leave an imprint on
radiation that intercepts a time-dependently ionised gas, like an
atmosphere affected by lightning discharges. In the case of brown
dwarfs, cyclotron maser emission is a good candidate for the 'probe
signal'. Observing the cyclotron periodic bursts for several
rotational periods (not necessarily consequent) is needed to spot a
variability. To interpret it as a signature of a flash ionisation
event which happened at a location along the cyclotron signal's path,
we suggest the following analysis of the detected varied
electromagnetic field:

\begin{enumerate}
	\item Several orders of magnitude's flash amplification of the
          initial signal, on a timescale comparable to the cyclotron
          radio emission wave-period, is a strong indication of a
          flash ionisation event. The detection of such amplitude
          amplification of the cyclotron emission (or any other signal
          carrier) requires a high time-resolution. The amplification
          seen in observations (Fig.~\ref{f:5}(a) for continuous
          emission, \ref{f:1}(b)-(c) for periodically seen emission)
          suggests an `intermediate' peak value of the conductivity
          flash produced by the ionisation being the cause.
	\item Flash (e.g., brief, short term, on a time-scale comparable to the cyclotron emission wave-period) damping of the observed radio signal (Fig.~\ref{f:5}(b) for continuous emission, \ref{f:1}(c)-(e) for periodically seen emission) can be a signature of a strong flash ionisation. It can though be confused with effects of beam variations and with spin-modulation of the cyclotron emission due to the source's rotation.
	\item Simultaneous observations in optical wavelength can help to distinguish the damping effect of the flash ionisation from effects of the beam variations (i.e., variations in the beam of electrons producing both optical aurora and radio emission). If there is no correlation between the radio variability and the variability in the optical aurora, the influence of beam variations can be ruled out as both radio and optical emissions are produced by the same beam of electrons. 
	\item Spin-modulation of the cyclotron radio emission due to
          the source's rotation \citep{morioka} can be ruled out or
          subtracted from the observations if observing the object for
          several consequent periods of rotation.
	\item Brown Dwarfs are thought to have turbulent atmospheres,
          as concluded from variability measurements and cloud
          detections (\citet{hell14}), e.g., variability seen in
          Luhman 16B~\citep{gillon13,burg14,luhman12}. The turbulent
          atmospheres have a higher rate of transient flash ionisation
          events like discharges or explosions, as seen for Saturn
          \citep{fischer11}. Some of Brown Dwarfs known for radio
          cyclotron emission are also fast rotators, e.g., TVLM513 has
          a period of just 1.9673~h \citep{doyle10,hall07}. This makes
          them easier objects for multi-period observations.
\end{enumerate}

\section{Summary of implications}\label{sec:4}

The motivation behind this research is to find a way to probe atmospheric discharge processes which are otherwise inaccessible. For this purposes, we investigated if and how a coherent emission (e.g. a cyclotron maser radio emission) could be modulated by passing through a region with flash ionisation processes in, for example, an atmospheric environment or a magnetosphere. Our results are applicable to a wider variety of problems in astrophysics when a pre-existing radiation travels through a rapidly ionising medium. These can include discharges in protoplanetary disks~\citep{muranushi12} and possibly fast radio bursts (see e.g.~\citep{thornton13} which could be attributed to flash transient events if their variations are not periodically detected\footnote{In the opposite case, when periodical field flashes are detected, those are more likely to be attributed to planets or stars companions to a pulsar~\citep{thornton13}.}. Our model (Eqs.~\eqref{e:16}--\eqref{e:18}) can also be applied to microwave background changed during the Reionisation Epoch~\citep{loeb01} if the limitations of 1D space-uniform consideration are reasonable for a particular scenario. 

We develop a model to describe the influence of flash ionisation events on pre-existing radiation that passes through the region where the transient events happen. We solved a (direct) problem of electromagnetic field transformation as a result of the flash transient events and provide a recipe for observers in Sect.~\ref{ssec:5.1}. Our main conclusions are
\begin{enumerate}
  \item The transformation of coherent emission by fast transient processes in the medium of its propagation can be observable;
  \item Attributing larger signal perturbations to more powerful processes is not always correct, as more powerful processes can lead to smaller response signals;
  \item Flash processes of intermediate intensity can be detected by short-term amplification of the signal. The response field's duration can be related to the flash ionisation's time scale and intensity (Fig.~\ref{f:7});
  \item If the initial emission signal is powerful enough to detect the character of the damping, the small pulse's amplitude and duration in the response field can provide information about the ionisation flash magnitude (peak value of conductivity) and duration;
  \item If the detection of the field's variation is possible for several related flashes (attributed to the same undergoing event, like multiple strokes of the same cloud-to-ground lightning discharge) and allows deriving the conductivity peak values from additional information, it would be possible to determine whether the discharges started as a runaway electrons breakdown or rather like capacity discharges, by deriving the dependence of the field responses' maximum values on maximum value of the conductivity change. Absence of a noticeable plateau in this dependence (see Fig.~\ref{f:6}) suggests that the runaway breakdown discharge is likely to be a source of the ionisation, which requires about ten times lower voltage to enable the discharge in comparison to a conventional capacity-like discharge. If the plateau is noticeable, then a capacitor-like discharge is more probable.
\end{enumerate}



\section*{Acknowledgments}

\bigskip
The authors gratefully acknowledge the support from the European
Community under the FP7 by the ERC starting grant~257431 and are
thankful for computer support from the University of St Andrews, SUPA
School of Physics \& Astronomy. Majority of reference search was done
with a help of ADS which is appreciated. We also thank Paul Rimmer for
reading and commenting on our manuscript.

\footnotesize{
\bibliographystyle{mn2e}
\bibliography{vorgul-3}{}
}


\end{document}